\documentclass{article}
\usepackage{amssymb}
\usepackage{amsmath}
\usepackage{enumerate}
\usepackage[dvips]{color}
\usepackage{graphicx}
\usepackage{tikz}
\usetikzlibrary{decorations.pathreplacing}
\usepackage{xcolor}
\usepackage{natbib}

\nonstopmode

\newcommand{\R }{\ensuremath{\mathbb R}}
\newcommand{\N }{\ensuremath{\mathbb N}}

\newcommand{\F} {\ensuremath{\mathcal{F}}}
\newcommand{\A} {\ensuremath{\mathcal{A}}}
\newcommand{\T} {\ensuremath{\mathbb{T}}}
\newcommand{\Prob} {\ensuremath{\mathbb{P}}}
\newcommand{\Q} {\ensuremath{\mathbb{Q}}}

\newcommand{\MM}{\mathcal{M}}
\newcommand{\MMF}{\mathcal{M}^f}

\newcommand{\FF}{\ensuremath{\mathbb{F}}}
\newcommand{\prF}{\ensuremath{\mathcal{F}^{{\rm pr}}}} 
\newcommand{\prFF}{\ensuremath{\mathbb{F}^{{\rm pr}}}} 
\newcommand{\prn}{\ensuremath{\mathcal{F}^{{\rm pr,n}}}}
\newcommand{\prN}{\ensuremath{\mathbb{F}^{{\rm pr,n}}}} 
\newcommand{\prk}{\ensuremath{\mathcal{F}^{{\rm pr,\ell}}}}
\newcommand{\prkp}{\ensuremath{\mathcal{F}^{{\rm pr,\ell+1}}}}

\newcommand{\pr}{\ensuremath{\Lambda}} 
\newtheorem{theorem}{Theorem}[section]

\newtheorem{corollary}[theorem]{Corollary}
\newtheorem{definition}[theorem]{Definition}
\newtheorem{example}[theorem]{Example}
\newtheorem{lemma}[theorem]{Lemma}

\newtheorem{proposition}[theorem]{Proposition}
\newtheorem{assumption}[theorem]{Assumption}
\newtheorem{remark}[theorem]{Remark}
\newenvironment{prova}[1][Proof]{\textbf{#1.} }{\ \rule{0.5em}{0.5em}}
\newcommand{\Lin}{\text{Lin}}

\begin{document}

\title{Pointwise Arbitrage Pricing Theory in Discrete Time}
\author{M. Burzoni, M. Frittelli, Z. Hou, M. Maggis and J. Ob\l \'oj}
\maketitle

\begin{abstract}
We develop a robust framework for pricing and hedging of
derivative securities in discrete-time financial markets. We
consider markets with both dynamically and statically traded
assets and make minimal measurability assumptions. We obtain an
abstract (pointwise) Fundamental Theorem of Asset Pricing and
Pricing--Hedging Duality. Our results are general and in
particular include so-called model independent results of
\cite{AB13,BFM16} as well as seminal results of \cite{DMW90} in a
classical probabilistic approach. Our analysis is scenario--based:
a model specification is equivalent to a choice of scenarios to be
considered. The choice can vary between all scenarios and the set
of scenarios charged by a given probability measure. In this way,
our framework interpolates between a model with universally
acceptable broad assumptions and a model based on a specific
probabilistic view of future asset dynamics.
\end{abstract}

\section{Introduction}

The \emph{State Preference Model} or \emph{Asset Pricing Model} underpins
most mathematical descriptions of Financial Markets. It postulates that the
price of $d$ financial assets is known at a certain initial time $t_{0}=0 $
(today), while the price at future times $t>0$ is unknown and is given by a
certain random outcome. To formalize such a model we only need to fix a
quadruple $(X ,\mathcal{F},\mathbb{F},S)$, where $X $ is the set of
scenarios, $\mathcal{F}$ a $\sigma $-algebra and $\mathbb{F}:=\{\mathcal{F}%
_{t}\}_{t\in I}\subseteq \mathcal{F}$ a filtration such that the $d$%
-dimensional process $S:=(S_{t})_{t\in I}$ is adapted. At this stage, no
probability measure is required to specify the Financial Market model $(X,%
\mathcal{F},\mathbb{F},S)$.\newline

One of the fundamental reasons for producing such models is to assign
rational prices to contracts which are not liquid enough to have a
market--determined price. Rationality here is understood via the economic
principle of absence of arbitrage opportunities, stating that it should not
be possible to trade in the market in a way to obtain a positive gain
without taking any risk. Starting from this premise, the theory of pricing
by no arbitrage has been successfully developed over the last 50 years. Its
cornerstone result, known as the Fundamental Theorem of Asset Pricing
(FTAP), establishes equivalence between absence of arbitrage and existence
of risk neutral pricing rules. The intuition for this equivalence can be
accredited to de Finetti for his work on \emph{coherence} and \emph{%
previsions} (see \citet{definetti,deF}). The first systematic attempt to understand the
absence of arbitrage opportunities in models of financial assets can be
found in the works of \citet{Ross76,Ross77} on capital pricing, see also
\citet{Hu82}. The intuition underpinning the arbitrage theory for derivative
pricing was developed by \citet{Samuelson:65}, \citet{BlackScholes:73} and
\citet{Merton:73}. The rigorous theory was then formalized by \citet{HK79} and extended in \citet{HP81}, see also
\citet{K81}. Their version of FTAP, in the case of a finite set of scenarios $%
X$, can be formulated as follows. Consider $X=\{\omega _{1},\ldots ,\omega
_{n}\}$ and let $s=(s^{1},\ldots ,s^{d})$ be the initial prices of $d$
assets with random outcome $S(\omega )=(S^{1}(\omega ),\ldots ,S^{d}(\omega
))$ for any $\omega \in X$. Then, we have the following equivalence
\begin{equation}\label{FTAPfinite}
\begin{array}{c}
\nexists H\in \mathbb{R}^{d}\text{ such that }H\cdot s\leq 0 \\
\text{ and }H\cdot S(\omega )\geq 0\text{ with $>$ for some }\omega \in X%
\end{array}%
\Longleftrightarrow
\begin{array}{c}
\exists Q\in \mathcal{P}\text{ such that }Q(\omega _{j})>0\text{ and } \\
E_{Q}[S^{i}]=s^{i},\forall \ 1\leq j\leq n,1\leq i\leq d%
\end{array}%
\end{equation}%
where $\mathcal{P}$ is the class of probability measures on $X$. In
particular, no reference probability measure is needed above and impossible
events are automatically excluded from the construction of the state space $X
$. On the other hand, linear pricing rules consistent with the observed
prices $s^{1},\ldots s^{d}$ and the No Arbitrage condition, turn out to be
(risk-neutral) probabilities with full support, that is, they assign
positive measure to any state of the world. By introducing a \emph{reference
probability measure} $P$ with full support and defining an arbitrage as a
portfolio with $H\cdot s\leq 0$, $P(H\cdot S(\omega )\geq 0)=1$ and $%
P(H\cdot S(\omega )>0)>0$, the thesis in \eqref{FTAPfinite} can be restated
as
\begin{equation}
\text{There is No Arbitrage}\Longleftrightarrow \exists Q\sim P\text{ such
that }E_{Q}[S^{i}]=s^{i}\quad \forall i=1,\ldots d.  \label{FTAPfinite2}
\end{equation}%
The identification suggested by \eqref{FTAPfinite2} allows non-trivial
extensions of the FTAP to the case of a general space $X$ with a fixed
reference probability measure, and was proven in
the celebrated work \citet{DMW90}, by use of measurable selection arguments.
It was then extended to continuous time models by \citet{DS,DS:FTAP}.\newline

The idea of introducing a reference probability measure to select scenarios
proved very fruitful in the case of a general $X$ and was instrumental for
the rapid growth of the modern financial industry. It was pioneered by
 \citet{Samuelson:65} and \citet{BlackScholes:73} who used it to formulate a
continuous time financial asset model with unique rational prices for all
contingent claims. Such models, with strong assumptions implying a unique
derivative pricing rule, are in stark contrast to a setting with little
assumptions, e.g.\ where the asset can follow any non-negative continuous
trajectory, which are consistent with many rational pricing rules. This
dichotomy was described and studied in the seminal paper of \citet{Merton:73}
who referred to the latter as ``assumptions sufficiently weak to gain
universal support\footnote{%
This setting has been often described as ``model--independent" but we see it
as a modelling choice with very weak assumptions.}" and pointed out that it
typically generates outputs which are not specific enough to be of practical
use. For that reason, it was the former approach, with the reference
probability measure interpreted as a probabilistic description of future
asset dynamics, which became the predominant paradigm in the field of
quantitative finance. The original simple models were extended, driven by
the need to capture additional features observed in the increasingly complex
market reality, including e.g.\ local or stochastic volatility. Such
extensions can be seen as enlarging the set of scenarios considered in the
model and usually led to plurality of rational prices. \newline

More recently, and in particular in the wake of the financial crisis, the
critique of using a single reference probability measure came from
considerations of the so-called Knightian uncertainty, going back to \citet{Knight:21},
and describing \emph{the model risk}, as contrasted with financial
risks captured within a given model. The resulting stream of research aims
at extending the probabilistic framework of \citet{DMW90} to a framework
which allows for a set of possible priors $\mathcal{R}\subseteq \mathcal{P}$%
. The class $\mathcal{R}$ represents a collection of plausible
(probabilistic) models for the market. In continuous time models this led
naturally to the theory of quasi-sure stochastic analysis as in \citet
{Denis2,Pe10,STZ11,STZ11a} and many further contributions, see e.g.\ \citet{DolinskySonerMOR}.
In discrete time, a general approach was
developed by \citet{BN13}. Under some technical conditions
on the state space and the set $\mathcal{R}$ they provide a version of the
FTAP, as well as the superhedging duality. Their framework includes the two
extreme cases: the classical case when $\mathcal{R}=\{P\}$ is a singleton
and, on the other extreme, the case of full ambiguity when $\mathcal{R}$
coincides with the whole set of probability measures and the description of
the model becomes pathwise. Their setup has been used to study a series of
related problems, see e.g.\ \citet{BayraktarZhang:16,ErhanZhou:16}. \bigskip

Describing models by specifying a family of probability measures $\mathcal{R}
$ appears natural when starting from the dominant paradigm where a reference
measure $P$ is fixed. However, it is not the only way, and possibly not the
simplest one, to specify a model. Indeed, in this paper, we develop a
different approach inspired by the original finite state space model used in
\citet{HP81} as well as the notion of \emph{prediction set} in \citet{Mykland_prediction_set}, see
also \citet{HO15}. Our analysis is scenario based. More specifically, \emph{%
agent's beliefs} or \emph{a model} are equivalent to selecting a set
of admissible scenarios which we denote by $\Omega \subseteq X$.
The selection may be formulated e.g.\ in terms of behaviour of
some market observable quantities and may reflect both the
information the agent has as well as the modelling assumptions she
is prepared to make. Our approach clearly includes the
\textquotedblleft universally acceptable" case of considering all
scenarios $\Omega =X$ but we also show that it subsumes the
probabilistic framework of \citet{DMW90}. Importantly, as we work
under minimal measurability requirement on $\Omega $, our models
offer a flexible way to interpolate between the two settings. The
scenario based specification of a model requires less
sophistication than selection of a family of probability measures
and appears particularly natural when considering (super)-hedging
which is a pathwise property.\newline

Our first main result, Theorem \ref{GFTAP}, establishes a Fundamental
Theorem of Asset Pricing for an arbitrary specification of a model $\Omega $
and gives equivalence between existence of a rational pricing rule (i.e.\ a
calibrated martingale measure) and absence of, suitably defined, arbitrage
opportunities. Interestingly the equivalence in \eqref{FTAPfinite} does not
simply extend to a general setting: specification of $\Omega $ which is
inconsistent with any rational pricing rule does not imply existence of
\emph{one arbitrage strategy}. Ex post, this is intuitive: while all agents
may agree that rational pricing is impossible they may well disagree on why
this is so. This discrepancy was first observed, and illustrated with an
example, by \citet{DH07}. The equivalence is only recovered
under strong assumptions, as shown by \citet{Riedel} in a topological
one-period setup and by \citet{AB13} in a general discrete
time setup. A rigorous analysis of this phenomenon in the case $\Omega =X$
was subsequently given by \citet{BFM16}, who also showed that several notions
of arbitrage can be studied within the same framework. Here, we extend their
result to an arbitrary $\Omega \subseteq X$ and to the setting with both
dynamically traded assets and statically traded assets. We show that also in
such cases agents' different views on arbitrage opportunities may be
aggregated in a canonical way into a pointwise arbitrage strategy in an
enlarged filtration. As special cases of our general FTAP, we recover
results in \citet{AB13,BFM16} as well as the classical
Dalang-Morton-Willinger theorem \citet{DMW90}. For the latter, we show that
choosing a probability measure $P$ on $X$ is equivalent to fixing a suitable
set of scenarios $\Omega ^{P}$ and our results then lead to probabilistic
notions of arbitrage as well as the probabilistic version of the Fundamental
Theorem of Asset pricing.\newline

Our second main result, Theorem \ref{thm: duality_options}, characterises
the range of rational prices for a contingent claim. Our setting is
comprehensive: we make no regularity assumptions on the model specification $%
\Omega $, on the payoffs of traded assets, both dynamic and static, or the
derivative which we want to price. We establish a pricing--hedging duality
result asserting that the infimum of prices of super-hedging strategies is
equal to the supremum of rational prices. As already observed in \citet
{BFM16b}, but also in \citet{BNT16} in the context of martingale optimal
transport, it may be necessary to consider superhedging on a smaller set of
scenarios than $\Omega $ in order to avoid a duality gap between rational
prices and superhedging prices. In this paper this feature is achieved
through the set of \emph{efficient} trajectories $\Omega _{\Phi }^{\ast }$
which only depends on $\Omega $ and the market. The set $\Omega _{\Phi
}^{\ast }$ recollects all scenarios which are supported by some rational
pricing rule. Its intrinsic and constructive characterisation is given in
the FTAP, Theorem \ref{GFTAP}. Our duality generalizes the results of \citet
{BFM16b} to the setting of abstract model specification $\Omega $ as well as
generic finite set of statically traded assets. The flexibility of model
choice is of particular importance, as stressed above. The \textquotedblleft
universally acceptable" setting $\Omega =X$ will typically produce wide
range of rational prices which may not be of practical relevance, as already
discussed by \citet{Merton:73}. However, as we shrink $\Omega $ from $X$ to a
set $\Omega ^{P}$, the range of rational prices shrinks accordingly and, in
case $\Omega ^{P}$ corresponds to a complete market model, the interval
reduces to a single point. This may be seen as a quantification of the
impact of modelling assumptions on rational prices and gives a powerful
description of model risk.\newline

We note that pricing--hedging duality results have a long history in the
field of robust pricing and hedging. First contributions focused on
obtaining explicit results working in a setting with one dynamically traded
risky asset and a strip of statically traded co-maturing call options with
all strikes. In his pioneering work \citet{Ho98} devised a methodology
based on Skorkohod embedding techniques and treated the case of lookback
options. His approach was then used in a series of works focusing on
different classes of exotic options, see \citet{BHR01,CO11,CO11b,CoxWang:11,HobsonKlimmek:12,HobsonNeuberger:12,HOST:14}.
More recently, it has been re-cast as an optimal transportation problem
along martingale dynamics and the focus shifted to establishing abstract
pricing--hedging duality, see \citet{BHLP13,DOR14,DS13,HO15}.\newline

The remainder of the paper is organised as follows. First, in Section \ref%
{mainresults}, we present all the main results. We give the necessary
definitions and in Section \ref{subsec:mainresults} state our two main
theorems: the Fundamental Theorem of Asset Pricing, Theorem \ref{GFTAP}, and
the pricing--hedging duality, Theorem \ref{thm: duality_options}, which we
also refer to as the superheding duality. In Section \ref{sect:ACC}, we
generalize the results of \citet{AB13} for a multi-dimensional non-canonical
stock process. Here, suitable continuity assumptions and presence of a
statically traded option $\phi _{0}$ with convex payoff with superlinear
growth allow to \textquotedblleft lift" superhedging from $\Omega _{\Phi
}^{\ast }$ to the whole $\Omega $. Finally, in Section \ref{sect:DMW}, we
recover the classical probabilistic results of \citet{DMW90}. The rest of the
paper then discusses the methodology and the proofs. Section \ref{sec:AA} is
devoted to the construction of strategy and filtration which aggregate
arbitrage opportunities seen by different agents. We first treat the case
without statically traded options when the so-called Arbitrage Aggregator is
obtained through a conditional backwards induction. Then, when statically
traded options are present, we devise a Pathspace Partition Scheme, which
iteratively identifies the class of polar sets with respect to calibrated
martingale measure. Section \ref{sec:proofs} contains the proofs with some
technical remarks relegated to the Appendix.

\section{Main Results}

\label{mainresults}

We work on a Polish space $X$ and denote $\mathcal{B}_{X}$ its Borel
sigma-algebra and $\mathcal{P}$ the set of all probability measures on $(X,%
\mathcal{B}_{X})$. If $\mathcal{G}\subseteq \mathcal{B}_{X}$ is a sigma
algebra and $P\in \mathcal{P}$, we denote with $\mathcal{N}^{P}(\mathcal{G}%
):=\{N\subseteq A\in \mathcal{G}\mid P(A)=0\}$ the class of $P$-null sets
from $\mathcal{G}$. 
We denote with $\F^\A$ the sigma-algebra generated by the analytic sets of $(X,\mathcal{B}_X)$ and with $\prF$ the sigma algebra generated by the class $\pr$
of projective sets of $(X,\mathcal{B}_{X})$. The latter is required for some of our technical arguments and we recall its properties in the Appendix. In particular, under a suitable choice of set theoretical axioms, it is included in the universal completion of $\mathcal{B}_X$, see Remark \ref{rk:projective_is_universal}.
As discussed in the introduction, we consider pointwise arguments and think of
a model as a choice of universe of scenarios $\Omega \subseteq X$. Throughout,
we assume that $\Omega$ is an analytic set.

\bigskip

Given a family of measures $\mathcal{R}\subseteq \mathcal{P}$ we say that a
set is polar (with respect to $\mathcal{R}$) if it belongs to $\left\{
N\subseteq A\in \mathcal{B}_{X}\ \mid \ Q(A)=0\ \forall Q\in \mathcal{R}%
\right\} $ and a property is said to hold \emph{quasi surely} ($\mathcal{R}$%
-q.s.) if it holds outside a polar set. For those random variables $g$ whose
positive and negative part is not $Q$-integrable ($Q\in \mathcal{P}$) we
adopt the convention $\infty -\infty =-\infty $ when we write $%
E_{Q}[g]=E_{Q}[g^{+}]-E_{Q}[g^{-}]$. Finally for any sigma-algebra $\mathcal{%
G} $ we shall denote by $\mathcal{L}(X,\mathcal{G} ;\R^d)$ the space of $%
\mathcal{G} $-measurable $d$-dimensional random vectors. For a given set $%
A\subseteq X$ and $f,g\in
\mathcal{L}(X,\mathcal{G} ;\R)$ we will often refer to $f\leq g$
on $A$ whenever $f(\omega)\leq g(\omega)$ for every $\omega\in A$
(similarly for $=$ and $<$).

\bigskip

We fix a time horizon $T\in \mathbb{N}$ and let $\T:=\left\{
0,1,...,T\right\} $. We assume the market includes both liquid
assets, which can be traded dynamically through time, and less
liquid assets which are only available for trading at time $t=0$.
The prices of  assets are represented by an
$\mathbb{R}^{d}$-valued stochastic process $S=(S_{t})_{t\in \T}$
on $(X,\mathcal{B}_{X})$. In addition we may also consider presence
of a vector of non-traded assets represented by an
$\mathbb{R}^{\tilde d}$-valued stochastic process $Y=(Y_{t})_{t\in
\T}$ on $(X,\mathcal{B}_{X})$ with $Y_0$ a constant, which may also be interpreted as
market factors, or additional information available to the agent.
The prices are given in units of some fixed numeraire asset
$S^{0}$, which itself is thus normalized: $S_{t}^{0}=1$ for all
$t\in \T$. In the presence of the additional factors $Y$, we let
$\mathbb{F}^{S,Y}:=(\mathcal{F}_{t}^{S,Y})_{t\in \T}$ be the
natural filtration generated by $S$ and $Y$ (When $Y\equiv 0$ we
have $\mathbb{F}^{S,0}=\mathbb{F}^{S}$ the natural filtration
generated by $S$). For technical reasons, we will also make use of
the filtration
$\prFF:=(\prF_t)_{t\in
\T}$ where $\prF_t$ is the sigma algebra
generated by the projective sets of
$(X,\mathcal{F}_{t}^{S,Y})$, namely  $\prF_t:=\sigma((S_u,Y_u)^{-1}(L)\mid \ L\in\pr,\ u\le t)$ (see the Appendix for further details). Clearly, $\mathbb{F}%
^{S,Y}\subseteq \prFF$ and $\prF_t$ is ``non-anticipative'' in the sense that the atoms of $\prF_t$ and $\mathcal{F}_{t}^{S,Y}$ are the same. Finally, we let $\Phi $
denote the vector of payoffs of the statically traded assets. We
consider the setting when $\Phi =\{\phi _{1},\ldots ,\phi _{k}\}$
is finite and each $\phi \in \Phi $ is
$\mathcal{F}^{\mathcal{A}}$-measurable. When there are no
statically traded assets we set $\Phi = 0$ which makes our
notation consistent.

\bigskip

For any filtration $\mathbb{F}$, $\mathcal{H}(\mathbb{F})$ is the
class of $\mathbb{F}$-predictable stochastic processes, with
values in $\mathbb{R}^{d}$, which represent admissible trading
strategies. Gains from investing in $S$, adopting a strategy $H$,
are given by
$(H\circ S)_{T}:=\sum_{t=1}^{T}\sum_{j=1}^{d}H_{t}^{j}(S_{t}^{j}-S_{t-1}^{j})=%
\sum_{t=1}^{T}H_{t}\cdot \Delta S_{t}$. In contrast, $\phi_j$ can
only be bought or sold at time $t=0$ (without loss of generality with zero initial cost)
and held until the maturity $T$, so that trading strategies are given by $\alpha\in \R^k$
and generate payoff $\alpha\cdot\Phi :=\sum_{j=1}^{k}\alpha_{j}\phi_{j}$ at time $T$.
We let $\A_\Phi(\FF)$ denote the set of such $\FF$-admissible trading strategies $(\alpha,H)$. 

\bigskip

Given a filtration $\mathbb{F}$, universe of scenarios $\Omega$ and set of statically traded assets $\Phi$, we let
\begin{eqnarray*}
\mathcal{M}_{\Omega,\Phi}(\mathbb{F})& := &\left\{ Q\in \mathcal{P}\mid S\text{ is an }%
\mathbb{F}\text{-martingale under }Q,\ Q(\Omega)=1\textrm{ and }E_{Q}[\phi]=0\;\forall \phi\in \Phi\right\}.
\end{eqnarray*}%
The support of a probability measure $Q$ is given by
$\text{supp}(P):=\bigcap \{C\in \mathcal{B}_X\mid C$ closed,
$P(C)=1\}$. We often consider measures with finite support and
denote it with a superscript ${}^f$, i.e.\ for a given set
$\mathcal{R}$ of probability measures we put $\mathcal{R}^f:=
\{Q\in \mathcal{R}\mid \text{supp}(Q)\text{ is finite}\}$. To wit,
$\mathcal{M}^{f}_{\Omega,\Phi }(\mathbb{F})$ denotes finitely
supported martingale measures on $\Omega$ which are calibrated to
options in $\Phi$. Define
\begin{equation}\label{filtr:options}
\mathbb{F}^{M}:=(\mathcal{F}^{M}_{t})_{t\in \T},\quad \textrm{ where } \mathcal{F}^{M}_{t}:=
\bigcap_{P\in \mathcal{M}_{\Omega,\Phi }(\mathbb{F}^{S,Y})}\mathcal{F}_{t}^{S,Y}\vee \mathcal{N}^{P}(\mathcal{F%
}_T^{S,Y}) \text{ ,}
\end{equation}
and we convene $\mathcal{F}^{M}_{t}$ is the power set whenever
$\mathcal{M}_{\Omega,\Phi }(\mathbb{F}^{S,Y})=\emptyset$.

\begin{remark}\label{rk:mart_measures}
In this paper we only consider filtrations $\mathbb{F}$ which
satisfy $\mathbb{F}^{S,Y}\subseteq \mathbb{F}\subseteq
\mathbb{F}^{M}$. All such filtrations generate the same set of
martingale measures, in the sense that  any $Q\in
\mathcal{M}_{\Omega,\Phi}(\mathbb{F})$ uniquely extends to a
measure $\hat{Q}\in \mathcal{M}_{\Omega,\Phi}(\mathbb{F}^{M})$
and, reciprocally, for any $\hat{Q}\in
\mathcal{M}_{\Omega,\Phi}(\mathbb{F}^{M})$, the restriction
$\hat{Q}_{\mid \mathbb{F}}$ belongs to
$\mathcal{M}_{\Omega,\Phi}(\mathbb{F})$. Accordingly, with a
slight abuse of
notation, we will write $\mathcal{M}_{\Omega,\Phi}(\mathbb{F})=%
\mathcal{M}_{\Omega,\Phi}(\mathbb{F}^{M})=\mathcal{M}_{\Omega,\Phi}$.
\end{remark}

In the subsequent analysis, the set of scenarios charged by martingale measures is crucial:
\begin{eqnarray}
\label{eq:OmegaPhi_def}
\Omega _{\Phi }^* & := & \left\{ \omega \in \Omega \mid \exists Q\in \mathcal{M}%
^f_{\Omega,\Phi}\text{ such that }Q(\omega )>0\right\} =\bigcup_{Q\in
\mathcal{M} ^f_{\Omega,\Phi}} supp(Q).
\end{eqnarray}%
We have by definition that for every $Q\in\mathcal{M}^f_{\Omega,\Phi }$
its support satisfies $\text{supp}(Q)\subseteq \Omega^*_{\Phi }$.
Notice that the key elements introduced so far namely $\mathcal{M}_{\Omega,\Phi}$,  $\mathcal{M}^f_{\Omega,\Phi}$, $\mathbb{F}^M$ and $\Omega^*_{\Phi}$, only depend on
the four basic ingredients of the market: $\Omega$, $S$, $Y$ and $\Phi$. Finally, in all of the above notations, we omit the subscript $\Omega$ when $\Omega=X$ and we omit the subscript $\Phi$ when $\Phi=0$, e.g.\ $\MMF$ denotes all finitely supported martingale measures on $X$.

\subsection{Fundamental Theorem of Asset Pricing and Superhedging Duality}
\label{subsec:mainresults}

We now introduce different notions of arbitrage opportunities
which play a key role in the statement of the pointwise Fundamental
Theorem of Asset Pricing.
\begin{definition}Fix a filtration $\mathbb{F}$, $\Omega\subseteq X$ and a set of statically traded options $\Phi$.
\begin{description}
\item[(1p)] A \emph{One-Point Arbitrage} (1p-Arbitrage) is a
strategy $(\alpha,H)\in \A_\Phi(\FF)$ such that $\alpha\cdot
\Phi+(H\circ S)_T\geq 0$ on $\Omega$ with a strict inequality for
some $\omega\in \Omega$.

\item[(SA)] A \emph{Strong Arbitrage} is a strategy
 $(\alpha,H)\in \A_\Phi(\FF)$ such that $\alpha\cdot \Phi+(H\circ S)_T>
0$ on $\Omega$.

\item[(USA)] A \emph{Uniformly Strong Arbitrage} is a strategy
$(\alpha,H)\in \A_\Phi(\FF)$ such that $\alpha\cdot \Phi+(H\circ
S)_T> \varepsilon$ on $\Omega$, for some $\varepsilon>0$.
\end{description}
\end{definition}

Clearly, the above notions are relative to the inputs and we often
stress this and refer to an arbitrage \emph{in} $\A_\Phi(\FF)$ and
\emph{on} $\Omega$. We are now ready to state the pathwise version
of Fundamental Theorem of Asset Pricing. It generalizes Theorem
1.3 in \citet{BFM16} in two directions: we include an analytic
selection of scenarios $\Omega$ and
we include static trading in options as well as dynamic trading in
$S$.

\begin{theorem}[Pointwise FTAP on $\Omega \subseteq X$]
\label{GFTAP} Fix $\Omega$ analytic and
$\Phi$ a finite set of $\mathcal{F}^{\mathcal{A}}$-measurable
statically traded options. Then, there exists a filtration
$\widetilde{\mathbb{F}}$ which aggregates arbitrage views in that:
\begin{equation*}
\text{No Strong Arbitrage in }\A_\Phi(\widetilde\FF) \text{ on
}\Omega \Longleftrightarrow \mathcal{M}_{\Omega,\Phi}(\mathbb{F}^{S,Y})\neq
\emptyset \Longleftrightarrow \Omega^*_\Phi\neq \emptyset
\end{equation*}
and $\mathbb{F}^{S,Y}\subseteq \widetilde{\mathbb{F}} \subseteq \mathbb{F}^M$.
Further, $\Omega^*_\Phi$ is analytic and
there exists a trading strategy $(\alpha^*,H^*)\in \A_\Phi(\widetilde\FF)$ which is an Arbitrage Aggregator in that $\alpha^*\cdot \Phi+(H^*\circ S)_T \geq 0$ on $\Omega$ and
\begin{equation}\label{def:OmegaPhi}
\Omega^*_{\Phi} = \left\{\omega\in\Omega \mid \alpha^*\cdot
\Phi(\omega)+(H^*\circ S)_T(\omega) = 0\right\}.
\end{equation}
Moreover, one may take $\widetilde\FF$ and $(\alpha^*,H^*)$ as constructed in \eqref{eq:Ftilde_options} and \eqref{Aggregator:options}  respectively.
\end{theorem}

We turn now to our second main result. For a given set of scenarios $A\subseteq X$,  define the
superhedging price on $A$:
\begin{equation}  \label{eq:sh_options}
\pi_{A,\Phi}(g):= \quad \inf \left\{ x\in \mathbb{R}\mid \exists
(\alpha,H)\in
\A_\Phi(\prFF) \text{ such that }x+\alpha\cdot\Phi+(H\circ
S)_{T}\geq g \text{ on } A\right\}.
\end{equation}
Following the intuition in \citet{BFM16b}, we expect to obtain
pricing--hedging duality only when considering superhedging on the
set of scenarios visited by martingales, i.e.\ we consider
$\pi_{\Omega_{\Phi}^*,\Phi}(g)$.

\begin{theorem}
\label{thm: duality_options} Fix $\Omega$ analytic and $\Phi$ a finite set of
$\mathcal{F}^{\mathcal{A}}$-measurable statically traded options.
Then, for any $\mathcal{F}^{\mathcal{A}}$-measurable $g$
\begin{equation}
\pi _{\Omega _{\Phi }^{\ast },\Phi }(g)=\sup_{Q\in\mathcal{M}_{\Omega ,\Phi }^{f}}E_{Q}[g]=\sup_{Q\in\mathcal{M}_{\Omega ,\Phi }}E_{Q}[g]
\label{eq:PHduality_options}
\end{equation}%
and, if finite, the left hand side is attained by some strategy
$(\alpha ,H)\in\A_\Phi(\prFF)$.
 \end{theorem}


The proofs of the two main theorems are given in Section
\ref{sec:proofs} below. We first prove the results when $\Phi=0$
and then extend iterating on the number of statically traded options
$k$. The proofs are intertwined and we explain their logic at the
beginning of Section \ref{sec:proofs}.
For the case with no options, it was claimed in \cite{BFM16b} that the superhedging strategy is universally measurable. Theorem \ref{thm: duality_options} corrects such statement, which remains true under the set-theoretic axioms that guarantee that projective sets are universally measurable (see Remark \ref{rk:projective_is_universal}).

The following proposition
is important as it shows that there are no One-Point arbitrage on
$\Omega$ if and only if each $\omega\in \Omega$ is weighted by
some martingale measure $Q\in \mathcal{M}^f_{\Omega,\Phi}$.

\begin{proposition}
\label{1p-Arb} Fix $\Omega$ analytic. Then
there are no One-Point Arbitrages on $\Omega$ with respect to
$\prFF $ if and only if $\Omega =
\Omega^*_{\Phi}$.
\end{proposition}

Under a mild assumption, this situation has further equivalent
characterisations:

\begin{remark}
 Under the additional
assumption that $\Phi$ is not perfectly replicable on $\Omega$,
the following are easily shown to be equivalent:

\begin{enumerate}
\item[(1)] No One-Point Arbitrage on $\Omega$ with respect to
$\prFF$.

\item[(2)] For any $x\in \R^k$, when $\varepsilon_x>0$ is small enough, $%
\mathcal{M}^f_{\Omega, \Phi+\varepsilon_x x} \neq \emptyset$.

\item[(3)] When $\varepsilon>0$ is small enough, for any $x\in \R^k$ such that $%
|x|< \varepsilon$, $\mathcal{M}^f_{\Omega, \Phi+x} \neq
\emptyset$.
\end{enumerate}

where $\Phi+x = \{\phi_1 + x_1, \ldots, \phi_n + x_n\}$.
\\In particular, small uniform modifications of the
statically traded options do not affect the existence of
calibrated martingale measures.
\end{remark}

\begin{remark}
If we choose all the probability measures on $\Omega$ as the reference class $\mathcal{P}$, then no One-Point Arbitrage corresponds to the notion of no-Arbitrage (NA($\mathcal{P}$)) considered in \citet{BN13}. The First Fundamental Theorem showed therein, applies to $\Omega$ equal to the T-fold product of a Polish space $\Omega_1$ (where T is the time horizon). Proposition \ref{1p-Arb} extends this result to $\Omega$ equal to an analytic subset of a general Polish space.
\end{remark}

\paragraph{Arbitrage de la classe $\mathcal{S}$.} In \citet{BFM16} a large variety of different notions of arbitrage were studied, with respect of a given class
of relevant measurable sets.

\begin{definition} Let $\mathcal{S}\subseteq \mathcal{B}_{X}$ be a class of measurable
subsets of $\Omega$ such that
$\emptyset \notin \mathcal{S}$. Fix a filtration $\mathbb{F}$ and
a set of statically traded options $\Phi$. An Arbitrage de la
classe $\mathcal{S}$ is a strategy $(\alpha,H)\in \A_\Phi(\FF)$
such that $\alpha\cdot \Phi+(H\circ S)_T\geq 0$ on $\Omega$ and
$\{\omega\in \Omega\mid \alpha\cdot \Phi+(H\circ S)_T > 0 \}$
contains a set in $\mathcal{S}$.
\end{definition}

We now apply our Theorem \ref{GFTAP} to characterize No Arbitrage
de la classe $\mathcal{S}$ in terms of the structure of the set of
martingale measures. In this way we generalize \citet{BFM16} to
the case of semi-static trading. Define $\mathcal{N}^M:=\{A\subseteq \Omega\mid\  Q(A)=0 \;\ \forall\, Q\in \mathcal{M}^{f}_{\Omega,\Phi}\}$.

\begin{corollary}[FTAP for the class $\mathcal{S}$]
\label{SFTAP} Fix $\Omega$ analytic and
$\Phi$ a finite set of $\mathcal{F}^{\mathcal{A}}$-measurable
statically traded options. Then, there exists a filtration
$\widetilde{\mathbb{F}}$ such that:
\begin{equation*}
\text{No Arbitrage del la classe $\mathcal{S}$ in
}\A_\Phi(\widetilde\FF) \text{ on }\Omega \Longleftrightarrow
\mathcal{N}^M\cap \mathcal{S}= \emptyset
\end{equation*}
and $\mathbb{F}^{S,Y}\subseteq \widetilde{\mathbb{F}} \subseteq
\mathbb{F}^M$.
\end{corollary}

\subsection{Pointwise FTAP for arbitrary many options in the spirit of
\citet{AB13}}\label{sect:ACC}

In this section, we want to recover and extend the main results in
\citet{AB13}. A similar result can be also found in \citet{CK16}
under slightly different assumptions. We work in the same setup as
above except that we can allow for a larger, possible uncountable,
set of statically traded options $\Phi=\{\phi_i: i \in I\}$.
Trading strategies $(\alpha,H)\in \A_\Phi(\prFF)$ correspond to
dynamic trading in $S$ using $H\in \mathcal{H}(\prFF)$
combined with a static position in a finite number of options in
$\Phi$.

\begin{assumption}\label{ass:options} In this section, we assume that $S$ takes values in $\R_+^{d\times (T+1)}$ and all the options $\phi\in \Phi$ are
continuous derivatives on the underlying assets $S$, more
precisely
$$\phi_i = g_i \circ S \quad \text{ for some continuous } g_i:\R_+^{d\times (T+1)}\to \R, \quad \forall i\in I.$$
In addition, we assume $0\in I$ and $\phi_0 = g_0(S_T)$ for a
strictly convex super-linear function $g_0$ on $\R^d$, such that other
options have a slower growth at infinity:
$$\hspace*{3cm}\lim_{|x|\to \infty}\frac{g_i(x)}{m(x)} = 0,  \quad \forall\ i \in  I/\{0\},\quad \textrm{where }m(x_0, . . . , x_T ) :=\sum_{t=0}^T g_0(x_t).$$
\\The option $\phi_0$ \textbf{can be only bought} at time $t=0$.
Therefore admissible trading strategies $\A_\Phi(\FF)$ consider
only positive values for the static position in $\phi_0$.
\end{assumption}

The presence of $\phi_0$ has the effect of restricting non-trivial
considerations to a compact set of values for $S$ and then the
continuity of $g_i$ allows to aggregate different arbitrages
without enlarging the filtration. This results in the following
special case of the pathwise Fundamental Theorem of Asset Pricing.
Denote by $\widetilde{\MM}_{\Omega,\Phi}:=\{Q\in
\MM_{\Omega,\Phi\setminus\{\phi_0\}} \mid E_Q[\phi_0]\leq  0\}$.
\begin{theorem}\label{thm: AB theorem 2}
Consider $\Omega$ analytic and such that  $\Omega =\Omega^*$, $\pi_{\Omega^*}(\phi_0) > 0$ and there exists
$\omega^*\in \Omega$ such that $S_0(\omega^*) = S_1(\omega^*) =\ldots=S_T(\omega^*)$. Under Assumption \ref{ass:options}, the following
are equivalent:
\begin{enumerate}
\item[(1)] There is no Uniformly Strong Arbitrage on $\Omega$ in
$\mathcal{A}_{\Phi}(\prFF)$; \item[(2)] There
is no Strong Arbitrage on $\Omega$ in
$\mathcal{A}_{\Phi}(\prFF)$; \item[(3)]
$\widetilde{\MM}_{\Omega,\Phi} \neq \emptyset.$
\end{enumerate}
Moreover, when any of these holds, for any upper semi-continuous
$g:\R^{d\times (T+1)}_+\to \R$  that satisfies
\begin{equation}\label{eq: payoff_dominatedA} \lim_{|x|\to
\infty}\frac{g^+(x)}{m(x)} = 0,
\end{equation}
the following pricing--hedging duality holds:
\begin{equation}
\pi_{\Omega, \Phi}(g(S))= \sup_{Q\in \widetilde{\MM}_{\Omega, \Phi}}
E_{Q}[g(S)]. \label{eq: main_thm 2}
\end{equation}
\end{theorem}


\begin{remark} We show in Remark \ref{noWayIfNoDominate} below that the pricing--hedging duality
may fail in general when super-replicating on the whole set
$\Omega$ as in \eqref{eq: main_thm 2}. This confirms the intuition
that the existence of an option $\phi_0$ which satisfies the
hypothesis of Theorem \ref{thm: AB theorem 2} is crucial. However
as shown in \citet{BFM16b} Section 4, the presence of such
$\phi_0$ is not sufficient. In fact the pricing hedging duality
\eqref{eq: main_thm 2} may fail if $g$ is not upper
semicontinuous.
\end{remark}

\subsection{Classical Model Specific setting and its selection of
scenarios}\label{sect:DMW}

In this section we are interested in the relation of our results
with the classical Dalang, Morton and Willinger approach from
\citet{DMW90}. For simplicity, and ease of comparison, throughout
this section we restrict to dynamic trading only: $\Phi=0$ and
$\A(\FF^\Prob)=\mathcal{H}(\FF^\Prob)$. For any filtration $\FF$,
we let $\FF^\Prob$ be the $\Prob$-completion of $\FF$. Recall that
a $(\FF,\Prob)$--arbitrage is a strategy $H\in \A(\FF^\Prob)$ such
that $(H\circ S)_T\geq 0$ $\Prob$-a.s.\ and $\Prob((H\circ
S)_T>0)>0$, which is the classical notion of arbitrage.

\begin{proposition} Consider a probability measure $\Prob\in \mathcal{P}$ and let $\mathcal{M}^{\ll\Prob}:=\{Q\in\mathcal{M}\mid Q\ll\Prob\}$.
There exists a set of scenarios $\Omega^{\Prob}\in
\mathcal{F}^{\mathcal{A}}$ and a filtration
$\widetilde{\mathbb{F}}$ such that $\mathbb{F}^{S,Y}\subseteq
\widetilde{\mathbb{F}} \subseteq \mathbb{F}^M$  and
\begin{equation*}
\text{No Strong Arbitrage in }\A(\widetilde{\mathbb{F}}) \text{ on
}\Omega^\Prob \Longleftrightarrow
\mathcal{M}^{\ll\Prob}\neq\emptyset .
\end{equation*}
Further,
\begin{equation*}
\text{No }(\FF^{S,Y},\Prob)\text{--arbitrage} \Longleftrightarrow
\Prob\left((\Omega^\Prob)^*\right)=1 \Longleftrightarrow
\mathcal{M}^{\sim\Prob}\neq\emptyset,
\end{equation*}
where $\mathcal{M}^{\sim\Prob}:=\{Q\in\mathcal{M}\mid Q\sim\Prob\}$.
\end{proposition}

\begin{prova}
For $1\leq t\leq T$, we denote $\chi_{t-1}$ the random set
$\chi_{\mathcal{G}}$ from \eqref{condSupp} with $\xi=\Delta S_t$
and $\mathcal{G}=\mathcal{F}^{S,Y}_{t-1}$ (see Appendix
\ref{conditional:support} for further details). Consider now the
set
$$U:=\bigcap_{t=1}^T\{\omega\in X \mid \Delta S_t(\omega) \in
\chi_{t-1}(\omega)\}$$ and note that, by Lemma \ref{probSupp} in the Appendix,
$U\in\mathcal{B}_X$ and $\Prob(U)=1$. Consider now the set $U^*$
defined as in \eqref{eq:OmegaPhi_def} (using $U$ in the place of
$\Omega$ and for $\Phi=0$) and define
\begin{equation*}
\Omega^\Prob:=\begin{cases}U&\text{ if }\Prob(U^*)>0\\U\setminus
U^*&\text{ if }\Prob(U^*)=0 \end{cases}\ ,
\end{equation*}
which satisfies $\Omega^\Prob\in\mathcal{F}^A$ and
$\Prob(\Omega^\Prob)=1$.

\bigskip

In both the proofs of sufficiency and necessity the existence of
the technical filtration is consequence of Theorem \ref{GFTAP}. To
prove sufficiency let $Q\in\mathcal{M}^{\ll\Prob}$ and observe
that, since $Q(\Omega^{\Prob})=1$, we have
$\mathcal{M}_{\Omega^\Prob}\neq \emptyset$. Since necessarily
$Q(U^*)>0$ we have $\Prob(U^*)>0$ and hence $\Omega^{\Prob}=U\in\mathcal{B}_X$. From Theorem \ref{GFTAP} we have No
$(\Omega^{\Prob},\tilde{\mathbb{F}})$ Strong Arbitrage. \\To prove
necessity observe first that $(\Omega^{\Prob})^*$ is either equal
to $U^*$, if $\Prob(U^*)>0$, or to the empty set otherwise. In the latter case, Theorem \ref{GFTAP} with $\Omega=U$ would contradict No $(\Omega^{\Prob},\tilde{\mathbb{F}})$
Strong Arbitrage. Thus, $(\Omega^\Prob)^*\neq\emptyset$ and $\Prob((\Omega^{\Prob})^*)=\Prob(U^*)>0$. Note
now that by considering $\tilde{\Prob}(\cdot):=\Prob(\cdot\mid
(\Omega^\Prob)^*)$ we have, by construction $0\in ri(\chi_{t-1})$
$\tilde{\Prob}-a.s.$ for every $1\leq t\leq T$, where $ri(\cdot)$
denotes the relative interior of a set. By \citet{Rokhlin} we
conclude that $\tilde{\Prob}$ admits an equivalent martingale
measure and hence the thesis. The last statement then also
follows.
\end{prova}


\section{Construction of the Arbitrage Aggregator and its filtration}
\label{sec:AA}

\subsection{The case without statically traded options}
\label{sec:AAnooptions}

The following Lemma is an empowered version of Lemma 4.4 in
\citet{BFM16}, which relies on measurable selections arguments,
instead of a pathwise explicit construction. In the following we
will set $\Delta S_{t}=S_{t}-S_{t-1}$ and $\Sigma _{t-1}^{\omega
}$ the level set of
the trajectory $\omega $ up to time $t-1$ of both traded and non-traded assets, i.e. \begin{equation}\label{level:set}\Sigma _{t-1}^{\omega }=\{%
\widetilde{\omega }\in X \mid S_{0:t-1}(\widetilde{\omega }
)=S_{0:t-1}(\omega )\text{ and } Y_{0:t-1}(\widetilde{\omega }
)=Y_{0:t-1}(\omega )\}, \end{equation} where
$S_{0:t-1}:=(S_0,\ldots,S_{t-1})$ and
$Y_{0:t-1}:=(Y_0,\ldots,Y_{t-1})$ . Moreover, by recalling that $\pr=\cup_{n\in\N}\Sigma^1_n$ (see the Appendix), we define $\prn_t:=\sigma((S_u,Y_u)^{-1}(L)\mid \ L\in\Sigma^1_n,\ u\le t)$.

\begin{lemma}
\label{conditional:splitting} Fix any $t\in \{1,\ldots ,T\}$ and $\Gamma \in
\pr$. There exist $n\in\N$, an index $\beta \in \{0,\ldots ,d\}$%
, random vectors $H^{1},\ldots ,H^{\beta }\in \mathcal{L}(X,\prn_{t-1};\mathbb{R}^{d})$, $\prn_{t}$%
-measurable sets $E^{0},...,E^{\beta }$ such that the sets $B^{i}:=E^{i}\cap
\Gamma ,$ $i=0,\ldots ,\beta $, form a partition of $\Gamma $\ satisfying:

\begin{enumerate}
\item \label{item_spezz_arb}if $\beta >0$ and $i=1,\ldots ,\beta $ then: $%
B^{i}\neq \emptyset $;$\ H^{i}\cdot \Delta S_{t}(\omega )>0$ for
all $\omega \in B^{i}$ and $H^{i}\cdot \Delta S_{t}(\omega )\geq
0$ for all $\omega \in \cup _{j=i}^{\beta }B^{j}\cup B^{0 }$.

\item \label{item_spezz}$\forall H\in \mathcal{L}(X,\prn_{t-1};\mathbb{R}^{d})$ such that $H\cdot \Delta S_{t}\geq 0$ on $B^{0}$ we
have $H\cdot \Delta S_{t}=0$ on $B^{0}$.
\end{enumerate}
\end{lemma}

\begin{remark}
\label{dep:t} Clearly if $\beta =0$ then $B^{0}=\Gamma $ (which
include the trivial case $\Gamma =\emptyset ).$ Notice also that
for any $\Gamma \in
\pr$ and $t=\{1,\dots ,T\}$ we have that $%
H^{i}=H_{t}^{i,\Gamma },B^{i}=B_{t}^{i,\Gamma },\beta =\beta _{t}^{\Gamma }$
depend explicitly on $t$ and $\Gamma $.
\end{remark}

\begin{prova}
Fix $t\in\{1,\dots,T\}$  and consider, for an arbitrary
$\Gamma\in \pr$, the multifunction
\begin{equation}\label{eq: multifunctionArb}
\psi _{t,\Gamma}:\omega\in X \mapsto \left\{ \Delta
S_{t}(\widetilde{\omega} )\mathbf{1}_{\Gamma}(\widetilde{\omega})
\mid \widetilde{\omega }\in \Sigma _{t-1}^{\omega }\right\}
\subseteq \mathbb{R}^{d}
\end{equation}%
where $\Sigma _{t-1}^{\omega }$ is defined in \eqref{level:set}. By definition of $\pr$, there exists $\ell\in\N$ such that $\Gamma\in\Sigma^1_\ell$. We first show that $%
\psi _{t,\Gamma}$ is an
$\prkp_{t-1}$-measurable multifunction. Note that for
any open set $O\subseteq \mathbb{R}^{d}$
\begin{equation*}
\{\omega \in X \mid \psi _{t,\Gamma}(\omega )\cap O\neq \emptyset
\}=(S_{0:t-1},Y_{0:t-1})^{-1}\left( (S_{0:t-1},Y_{0:t-1})\left(
B\right) \right),
\end{equation*}
where $B= (\Delta S_{t}\mathbf{1}_{\Gamma})
^{-1}(O)$. First $\Delta S_{t}\mathbf{1}_{\Gamma}$ is an $\prk$%
-measurable random vector then $B\in \prk$, the sigma-algebra generated by the $\ell$-projective sets of $X$. Second $%
S_{u}, Y_u$ are Borel measurable functions for any $0\leq u\leq
t-1$ so that, from Lemma \ref{Fsigma}, we have that
$(S_{0:t-1},Y_{0:t-1})(B)$ belongs to the sigma-algebra generated
by the ($\ell+1$)-projective sets of $Mat((d+\tilde{d})\times t;\mathbb{R})$ (the space
of $(d+\tilde{d})\times t$ matrices with real entries) endowed with its
Borel sigma-algebra. Applying again Lemma \ref{Fsigma} we deduce
that $(S_{0:t-1},Y_{0:t-1})^{-1}\left( (S_{0:t-1},Y_{0:t-1})\left(
B\right)\right)\in \prkp_{t-1}$ and hence the
desired measurability for $\psi _{t,\Gamma}$.\newline Let
$\mathbb{S}^{d}$ be the unit sphere in $\R^d$, by preservation of
measurability (see \citet{R}, Chapter 14-B) the following
multifunction is closed valued and
$\prkp_{t-1}$-measurable
\begin{equation*}
\psi _{t,\Gamma}^{\ast }(\omega ):=\left\{ H\in \mathbb{S}^{d}\mid
H\cdot y\geq 0\quad \forall y\in \psi _{t,\Gamma}(\omega
)\right\}.
\end{equation*}
It follows that it admits a Castaing representation (see
Theorem 14.5 in \citet{R}), that is, there exists a countable
collection of measurable functions
$\{\xi^n_{t,\Gamma}\}_{n\in\mathbb{N}}\subseteq
\mathcal{L}(X,\prkp_{t-1};\R^d)$ such that
$\psi
_{t,\Gamma}^{\ast }(\omega)=\overline {\{\xi_{t,\Gamma}^n(\omega)\mid n\in%
\mathbb{N}\}}$ for every $\omega$ such that $\psi
_{t,\Gamma}^{\ast }(\omega )\neq \emptyset $ and
$\xi^n_{t,\Gamma}(\omega)=0$ for every $\omega$ such that $\psi
_{t,\Gamma}^{\ast }(\omega )= \emptyset $  . Recall that every
$\xi_{t,\Gamma}^n$
is a measurable selector of $\psi _{t,\Gamma}^{\ast }$ and hence, $%
\xi_{t,\Gamma}^n\cdot\Delta S_t\geq 0$ on $\Gamma$. Note moreover
that,
\begin{equation}  \label{maxPos}
\forall\,\omega\in X, \quad \bigcup_{\xi\in\psi _{t,\Gamma}^{\ast
}(\omega)}\left\{y\in\mathbb{R}^d\mid \xi\cdot y>0\right\}=\bigcup_{n\in%
\mathbb{N}}\left\{y\in\mathbb{R}^d\mid \xi_{t,\Gamma}^n(\omega)\cdot
y>0\right\}
\end{equation}
The inclusion $(\supseteq)$ is clear, for the converse note that if $y$
satisfies $\xi_{t,\Gamma}^n(\omega)\cdot y\leq 0$ for every $n\in\mathbb{N}$
then by continuity $\xi\cdot y\leq 0$ for every $\xi\in\psi
_{t,\Gamma}^{\ast }(\omega)$.  \newline
We now define the the conditional standard separator as

\begin{equation}  \label{standard:separator}
\xi_{t,\Gamma}:=\sum_{n=1}^{\infty} \frac{1}{2^n} \xi_{t,\Gamma}^n
\end{equation}

which is $\prkp_{t-1}$-measurable and, from %
\eqref{maxPos}, satisfies the following maximality property: $\{\omega\in X
\mid \xi(\omega)\cdot\Delta S_{t}(\omega)>0\}\subseteq \{\omega\in X\mid
\xi_{t,\Gamma}(\omega)\cdot\Delta S_{t}(\omega)>0\}$ for any $\xi$
measurable selector of $\psi _{t,\Gamma}^{\ast }$.

\begin{description}
\item[Step 0:] We take $A^{0}:=\Gamma $ and consider the
multifunction $\psi _{t,A^{0}}^{\ast }$ and the conditional
standard separator $\xi _{t,A^{0}}$ in \eqref{standard:separator}.
If $\psi _{t,A^{0}}^{\ast }(\omega )$ is a linear subspace of
$\R^{d}$ (i.e. $H\in \psi _{t,A^{0}}^{\ast }(\omega )$
implies necessarily $-H\in \psi _{t,A^{0}}^{\ast }(\omega )$ ) for any $%
\omega \in A^{0}$ then set $\beta =0$ and $A^{0}=B^{0}$ (In this
case obviously $E^0=X$).

\item[Step 1:] If there exists an $\omega \in A^{0}$ such that
$\psi _{t,A^{0}}^{\ast }(\omega )$ is not a linear subspace of
$\R^{d}$ then we set $H_{1}=\xi _{t,A^{0}}$, $E^{1}=\{\omega \in X
\mid H_{1}\Delta S_{t}>0\}$, $B^{1}=\{\omega \in A^{0}\mid
H_{1}\Delta S_{t}>0\}=E^1\cap \Gamma$ and $A^{1}=A^{0}\setminus
B^{1}=\{\omega \in A^{0}\mid H_{1}\Delta S_{t}=0\}$. If now $\psi
_{t,A^{1}}^{\ast }(\omega )$ is a linear subspace
of $\R^{d}$ for any $\omega \in A^{1}$ then we set $\beta =1$ and $%
A^{1}=B^{0}$. If this is not the case we proceed iterating this scheme.

\item[Step 2:] notice that for every $\omega \in A^{1}$ we have
$\Delta S_{t}(\omega )\in R_{1}(\omega ):=\{y\in
\mathbb{R}^{d}\mid H_{1}(\omega )\cdot y=0\}$ which can be
embedded in a subspace of $\mathbb{R}^{d}$ whose dimension is
$d-1$. We consider the case in which there exists one $\omega \in
A^{1}$ such that $\psi _{t,A^{1}}^{\ast }(\omega )$ is not a
linear subspace of $R_{1}(\omega )$: we set $H_{2}=\xi
_{t,A^{1}}$, $E^{2}=\{\omega \in X\mid H_{2}\Delta S_{t}>0\}$,
$B^{2}=\{\omega \in A^{0}\mid H_{2}\Delta S_{t}>0\}=E^2\cap
\Gamma$ and $A^{2}=A^{1}\setminus B^{2}=\{\omega \in A^{1}\mid
H_{2}\Delta S_{t}=0\}$. If now $\psi _{t,A^{2}}^{\ast }(\omega )$
is a linear subspace of $R_{1}(\omega )$ for any $\omega \in
A^{2}$ then we set $\beta =2$ and $A^{2}=B^{0}$. If this is not
the case we proceed iterating this scheme.
\end{description}

The scheme can be iterated and ends at most within $d$ Steps, so that, there exists $n\le \ell+2d$ yielding the desired measurability.
\end{prova}


Define, for $\Omega\in\pr$,
\begin{eqnarray}
\Omega_{T}:= && \Omega \nonumber \\
\Omega _{t-1}:= &&\Omega _{t}\setminus \bigcup_{i=1}^{\beta_t} B^i_t,\text{%
\quad }t\in \{1,\ldots,T\}, \label{Omega_t}
\end{eqnarray}%
where $B^i_t:= B^{i,\Gamma}_t$, $\beta_t:=\beta_t^{\Gamma}$ are the sets and
index constructed in Lemma \ref{conditional:splitting} with $\Gamma=\Omega_t$%
, for $1\leq t\leq T$. Note that we can iteratively apply Lemma \ref{conditional:splitting} at time $t-1$ since $\Gamma=\Omega_{t}\in\pr$.

\begin{corollary}\label{polars}
For any $t\in\{1,\ldots,T\}$, $\Omega$ analytic and $Q\in%
\mathcal{M}_{\Omega}$ we have $\cup_{i=1}^{\beta_t} B^i_t$ is a subset of a $%
Q$-nullset. In particular $\cup_{i=1}^{\beta_t} B^i_t$ is an $\mathcal{M}_{\Omega}$ polar set.
\end{corollary}

\begin{prova}
Let $\Gamma=\Omega$. First observe that the map $\psi _{T,\Gamma}$ in  \eqref{eq: multifunctionArb} is $\F^\A_{T-1}$-measurable. Indeed the set $B= (\Delta S_{t}\mathbf{1}_{\Gamma})
^{-1}(O)$ is analytic since it is equal to $\Delta S_{t}^{-1}(O)\cap \Gamma$ if $0\notin O$ or $\Delta S_{t}^{-1}(O)\cup \Gamma$ if $0\in O$. The measurability of $\psi _{T,\Gamma}$ follows from Lemma \ref{lem:img}. As a consequence, $H^1$ and $B^1$ from Lemma \ref{conditional:splitting} satisfy: $H^1\in\mathcal{L}(X,\F^\A_{T-1};\mathbb{R}^{d})$ and $B^1=\{H^1\cdot\Delta S_T>0\}\in\F^\A$. Suppose $Q(B^1)>0$. The strategy $H_u:=H^1\mathbf{1}_{T-1}(u)$ satisfies:
\begin{itemize}
\item $H$ is $\mathbb{F}^Q$-predictable, where $\mathbb{F}^Q=\{\mathcal{F}
_t^S\vee \mathcal{N}^Q(\mathcal{B}_X)\}_{t\in\{0,\dots,T\}}$.

\item $(H\cdot S)_T\geq 0$ $Q$-a.s. and $(H\cdot S)_T>0$ on
$B^1$ which has positive probability.
\end{itemize}
Thus, $H$ is an arbitrage in the classical probabilistic sense, which leads
to a contradiction. Since $B^1$ is a $Q$-nullset, there exists $\tilde{B}^1\in\mathcal{B}_X$ such that $B^1\subseteq\tilde{B}^1$ and $Q(\tilde{B}^1)=0$. Consider now the Borel-measurable version of $S_T$ given by $\tilde{S}_T=S_T\mathbf{1}_{X\setminus\tilde{B}^1}+S_{T-1}\mathbf{1}_{\tilde{B}^1}$. We iterate the above procedure replacing $S$ with  $\tilde{S}$ at each step up to time $t$. As in Lemma \ref{conditional:splitting}, the procedure ends in a finite number of step yielding a collection $\{\tilde{B}^i_t\}_{i=1}^{\tilde{\beta}_t}$ such that $\cup_{i=1}^{\beta_t} B^i_t\subseteq \cup_{i=1}^{\tilde{\beta}_t} \tilde{B}^i_t$ with $Q(\cup_{i=1}^{\beta_t} \tilde{B}^i_t)=0$.

\end{prova}

\begin{corollary}\label{one:step:martingale}
Let $B_t^0$ the set provided by Lemma
\ref{conditional:splitting} for $\Gamma =\Omega _{t}$. For every
$\omega \in B_t^0$ there exists
$Q\in \mathcal{P}^{f}$ with $Q(\{\omega \})>0$ such that $\mathbb{E}%
_{Q}[S_{t}\mid \mathcal{F}_{t-1}^{S}](\omega)=S_{t-1}(\omega)$.
\end{corollary}

\begin{prova}
Fix $\omega \in B_{t}^{0}$ and let $\Sigma _{t-1}^{\omega }$ be
given as in \eqref{level:set}. We consider $D:=\Delta
{S}_{t}(\Sigma _{t-1}^{\omega }\cap B_t^0)\subseteq
\R^{d}$ and $C:=\{\lambda v\mid v\in conv(D),\ \lambda \in
\R^{+}\}$ where $conv(D)$ denotes the convex hull of $D$. Denote
by $ri(C)$ the relative interior of $C
$. From Lemma \ref{conditional:splitting} item \ref{item_spezz} we have $%
H\cdot \Delta {S}_{t}(\widetilde{\omega })\geq 0$ for all $\widetilde{\omega
}\in \Sigma _{t-1}^{\omega }\cap B_t^0$ implies $H\cdot \Delta {S}%
_{t}(\widetilde{\omega })=0$ for all $\widetilde{\omega }\in \Sigma
_{t-1}^{\omega }\cap B_t^0$, which is equivalent to $0\in ri(C)$.
From Remark 4.8 in \citet{BFM16} we have that for every $x\in D$ there exists
a finite collection $\{x_{j}\}_{j=1}^{m}\subseteq D$ and $\{\lambda
_{j}\}_{j=1}^{m+1}$ with $0<\lambda _{j}\leq 1$, $\sum_{j=1}^{m+1}\lambda
_{j}=1$, such that
\begin{equation}
0=\sum_{j=1}^{m}\lambda _{j}x_{j}+\lambda _{m+1}x.  \label{convexComb}
\end{equation}%
Choose now $x:=\Delta S_{t}(\omega )$ and note that for every $j=1,\ldots m$
there exists $\omega _{j}\in \Sigma _{t-1}^{\omega }\cap B_t^0$ such
that $\Delta {S}_{t}(\omega _{j})=x_{j}$. Choose now $Q\in \mathcal{P}^{f}$
with conditional probability $Q(\cdot \mid \mathcal{F}_{t-1}^{S})(\omega):=%
\sum_{j=1}^{m}\lambda _{j}\delta _{\omega _{j}}+\lambda _{m+1}\delta
_{\omega }$, where $\delta _{\widetilde{\omega }}$ denotes the Dirac measure
with mass point in $\widetilde{\omega }$. From \eqref{convexComb}, we have
the thesis.
\end{prova}

\begin{lemma}\label{LemNOpolar}For $\Omega\in\pr$, the set $\Omega^*$, defined in \eqref{eq:OmegaPhi_def} with $\Phi=0$, coincides with $\Omega_0$ defined in \eqref{Omega_t}, and therefore
$\Omega^*\in\pr$. Moreover, if $\Omega$ is analytic then $\Omega^{*}$ is analytic and we have the
following
\begin{equation*}
\Omega^{*}\neq \emptyset \Longleftrightarrow \mathcal{M}_{\Omega}
\neq \emptyset
\Longleftrightarrow \mathcal{M}_{\Omega}^{f}\neq \emptyset .
\end{equation*}
\end{lemma}

\begin{prova} The proof is analogous to that of Proposition 4.18 in  \citet{BFM16},
but we give here a self-contained argument. Notice that
$\Omega^*\subseteq \Omega_0$ follows from the definitions and Corollary \ref{polars}.
For the reverse inclusion, it suffices to show that for $\omega_{\ast}\in \Omega_0$ there
exists a $Q\in \mathcal{M}^{f}_{\Omega}$ such that $Q(\{\omega_{\ast
}\})>0$, i.e. $\omega_{\ast}\in \Omega^*$. From Corollary
\ref{one:step:martingale}, for any $1\leq t\leq T$, there exists a finite
number of elements of $\Sigma _{t-1}^{\omega }\cap B_t^0$
named $C_{t}(\omega ):=\{\omega ,\omega _{1},\ldots ,\omega
_{m}\}$, such that
\begin{equation}
S_{t-1}(\omega )=\lambda _{t}(\omega )S_{t}(\omega
)+\sum_{j=1}^{m}\lambda _{t}(\omega _{j})S_{t}(\omega _{j})
\label{first}
\end{equation}%
where $\lambda _{t}(\omega )>0$ and $\lambda _{t}(\omega
)+\sum_{j=1}^{m}\lambda _{t}(\omega _{j})=1$.\newline
Fix now $\omega _{\ast }\in \Omega _{0}$. We iteratively build a set $%
\Omega _{f}^{T}$ which is suitable for being the finite support of
a discrete martingale measure (and contains $\omega _{\ast
}$).\newline
Start with $\Omega _{f}^{1}=C_{1}(\omega _{\ast })$ which satisfies (\ref%
{first}) for $t=1$. For any $t>1$, given $\Omega _{f}^{t-1}$,
define $\Omega _{f}^{t}:=\left\{ C_{t}(\omega )\mid \omega \in
\Omega _{f}^{t-1}\right\} $. Once $\Omega _{f}^{T}$ is settled, it
is easy to construct a martingale measure via \eqref{first}:
\begin{equation*}
Q(\{\omega \})=\prod_{t=1}^{T}\lambda _{t}(\omega )\quad \forall
\omega \in \Omega _{f}^{T}
\end{equation*}%
Since, by construction, $\lambda _{t}(\omega _{\ast })>0$ for any
$1\leq t\leq T $, we have $Q(\{\omega _{\ast }\})>0$ and $Q\in
\mathcal{M}_{\Omega}^f$.\\ For the last assertion, suppose $\Omega$ is analytic. From Remark 5.6 in \citet{BFM16b}, $\Omega^*$ is also analytic. In particular, if $\mathcal{M}_{\Omega}\neq\emptyset$ then, from Corollary \ref{polars}, $Q(\Omega^*)=1$ for any $Q\in\mathcal{M}_{\Omega}$. This implies $\Omega^*\neq\emptyset$. The converse implication is trivial.
\end{prova}

\begin{lemma}\label{NoOnePointNoOptions} Suppose $\Phi=0$. Then, no One-Point Arbitrage $\Leftrightarrow$ $\Omega^*=\Omega$.
\end{lemma}
\begin{prova}
``$\Leftarrow$" If $(H\circ S)_T\geq 0$ on $\Omega$ then  $(H\circ S)_T= 0$ $Q$-a.s. for every $Q\in\mathcal{M}_{\Omega}^f$. From the hypothesis we have $\cup\{\text{supp}(Q)\mid Q\in\mathcal{M}_{\Omega}^f\}=\Omega$ from which the thesis follows.
``$\Rightarrow$". Let $1\leq t\leq T$ and $\Gamma=\Omega_t$.
Note that if $\beta_t$ from Lemma \ref{conditional:splitting} is strictly positive then $H^1$ is a One-Point Arbitrage. We thus have $\beta_t=0$ for any $1\leq t\leq T$ and hence $\Omega_0=\Omega$. From Lemma \ref{LemNOpolar} we have $\Omega^*=\Omega$.
\end{prova}

\begin{definition}
\label{defUArb} We call Arbitrage Aggregator
the process
\begin{equation}
H_{t}^*(\omega ):= \sum_{i=1}^{\beta_t} H_t^{i,\Omega_t}(\omega)\mathbf{1}%
_{B_t^{i,\Omega_t}}(\omega)  \label{universal}
\end{equation}%
for $t\in \left\{ 1,\ldots,T\right\}$, where $H_t^{i,\Omega_t},
B_t^{i,\Omega_t}, \beta_t$ are provided by Lemma \ref{conditional:splitting} with $\Gamma=\Omega_t$.
\end{definition}

\begin{remark}
\label{strictlyPos}  Observe that from Lemma
\ref{conditional:splitting}
item \ref{item_spezz_arb},  $(H^*\circ S)_T(\omega)\geq 0$ for all $%
\omega\in\Omega$ and from Lemma \ref{LemNOpolar}, $(H^*\circ S)_T(\omega)>0$ for all $\omega\in\Omega\setminus\Omega^*$%
.
\end{remark}

\begin{remark}
By construction we have that $H_t^*$ is $\prn$%
-measurable for every $t\in\{1,\dots,T\}$, for some $n\in\N$. Moreover any
$B_t^{i,\Omega_t}$ is the intersection of an $\prn
_t$-measurable set with $\Omega_t$. As a consequence
we have $(H_t^*)_{\mid \Omega_t}:\Omega_t\rightarrow \R^d$ is
$(\prn _t)_{\mid \Omega_t}$-measurable.
\end{remark}

\begin{remark}\label{technical:0options} In case there are no options to be statically
traded, $\Phi=0$, the enlarged filtration $\widetilde{\mathbb{F}}$ required in Theorem \ref{GFTAP} is
given by
\begin{eqnarray}
\widetilde{\mathcal{F}}_{t} &:&=\mathcal{F}^{S,Y}_t\vee\sigma
(H^*_{1},\ldots,H^*_{t+1}),\text{ }t\in
\left\{0,\ldots,T-1\right\}
\label{enlargment1} \\
\widetilde{\mathcal{F}}_{T} &:&=\mathcal{F}^{S,Y}_T\vee\sigma
(H_{1}^*,\ldots,H^*_{T}).  \label{enlargment2}
\end{eqnarray}
so that the Arbitrage Aggregator from \eqref{universal} is
predictable with respect to $\widetilde{\mathbb{F}}
=\{\widetilde{\mathcal{F}}_{t}\}_{t\in \T}$.
\end{remark}

\subsection{The case with a finite number of statically traded options}
\label{partition:scheme}

Throughout this section we consider the case of a finite set of
options $\Phi$. As in the previous section we consider $\prn_t:=\sigma((S_u,Y_u)^{-1}(L)\mid \ L\in\Sigma^1_n,\ u\le t)$ and $\prN:=(\prn_t)_{t\in\T}$.

\begin{definition}
\label{def:pps} A pathspace partition scheme $\mathcal{R}(\alpha^{%
\star},H^{\star})$ of $\Omega$ is a collection of trading strategies $%
H^1,\ldots, H^{\beta}\in \mathcal{H}(\prN)$, for some $n\in\N$, $%
\alpha^1, \ldots, \alpha^{\beta}\in \R^k$ and arbitrage
aggregators $\tilde{H}^0, \ldots, \tilde{H}^{\beta}$, for some
$1\leq \beta\leq k$, such that

\begin{itemize}
\item[(i)] $\alpha^i$, $1\leq i\leq \beta$, are linearly
independent,

\item[(ii)] for any $i\le \beta$,
\begin{equation*}
(H^i \circ S)_T + \alpha^i \cdot \Phi \ge 0 \text{ on $A^*_{i-1}$,
}
\end{equation*}
where $A_0 = \Omega$, $A_{i} := \{ (H^i \circ S)_T + \alpha^i
\cdot \Phi = 0 \} \cap A^*_{i-1}$ and $A^*_{i}$ is the set
$\Omega^*$ in \eqref{eq:OmegaPhi_def} with $\Omega=A_{i}$ and $\Phi=0$ for
$1\le i\le\beta$,

\item[(iii)] for any $i = 0, \ldots, \beta$, $\tilde{H}^i$ is the
Arbitrage Aggregator, as defined in \eqref{universal} substituting
$\Omega$ with $A_{i}$,

\item[(iv)] if $\beta < k$, then either $A^*_{\beta} = \emptyset$,
or for any $\alpha\in \R^k$ linearly independent from $\alpha^1,
\ldots, \alpha^{\beta}$, there does not exist $H$ such that
\begin{equation*}
(H \circ S)_T + \alpha \cdot \Phi \ge 0 \text{ on $A^*_{\beta}$. }
\end{equation*}
\end{itemize}
\end{definition}

We note that as defined in (ii) above, each $A_i\in \pr$ so that $A_i^*\in \pr$ by Lemma \ref{LemNOpolar}. The purpose of a pathspace partition scheme is to iteratively split the pathspace $\Omega$ in subsets on which a {%
Strong Arbitrage strategy can be identified. For the existence of calibrated martingale measure it will be crucial to see whether this procedure exhausts the pathspace or not.
Note that on $A_i$ we can perfectly replicate $i$ linearly independent combinations of options $\alpha^j\cdot \Phi$, $1\leq j\leq i$. In consequence, we make at most $k$ such iterations, $\beta\leq k$, and if $\beta=k$ then all statically traded options are perfectly replicated on $A^*_{\beta}$ which reduces here to the setting without statically traded options.

\begin{definition}
A pathspace partition scheme
$\mathcal{R}(\alpha^{\star},H^{\star})$ is successful if
$A^*_\beta\neq \emptyset$.
\end{definition}

We illustrate now the construction of a successful pathspace partition scheme.
\begin{example}\label{ex: partition}
 Let $X=\mathbb{R}^2$. Consider a financial market with one dynamically traded asset
$S$ and two options available for static trading
$\Phi:=(\phi_1,\phi_2)$. Let $S$ be the canonical process i.e.
$S_t(x)=x_t$ for $t=1,2$ with initial price $S_0=2$. Moreover, let
$\phi_i:=g_i(S_1,S_2)-c$ for $i=1,2$ with $c>0$,
$g_i:=(x_2-K_i)^+\mathbf{1}_{[0,b]}(x_1)+c\mathbf{1}_{(b,\infty)}(x_1)$ and $K_1>K_2$. Namely,
$\phi_i$ is a knock in call option on $S$ with maturity $T=2$,
strike price $K_i$, knock in value $b\geq 0$ and cost $c>0$, but the cost is recovered if the option is not knocked in. Consider the pathspace selection $\Omega=[0,4]\times[0,4]$.
\end{example}
Start with $A_0:=\Omega$ and suppose $0<K_2<b<2$.
\begin{enumerate}
\item the process
$(\tilde{H}^0_1,\tilde{H}^0_2):=(0,\mathbf{1}_{\{0\}}(S_1)-\mathbf{1}_{\{4\}}(S_1))$
is an arbitrage aggregator: when $S_1$ hits the values $\{0,4\}$,
the price process does not decrease, or increase, respectively. It
is easily seen that there are no more arbitrages on $A_0$ from
dynamic trading only. Thus, $ A_0^*=\{(0,4)\times
[0,4]\}\cup\{\{0\}\times \{0\}\}\cup\{\{4\}\times \{4\}\}$. \item
Suppose now we have a semi-static strategy $(H^1,\alpha^1)$ such
that
\begin{equation*}
(H^1\circ S)_T+\alpha^1\cdot \Phi\geq 0\text{ on }A_0^*.
\end{equation*}
where $\alpha^1=(-\alpha,\alpha)\in \R^2$ for some $\alpha>0$
(since $K_1>K_2$ and $\phi_{1}, \phi_2$ have the same cost).
Moreover since $\Phi=0$ if $S_1\notin[0,b]$, we can choose
$(H^1_1,H^1_2)=(0,0)$. A positive gain is obtained on
$B_1:=[0,b]\times (K_2,4]$ (see Figure \ref{FigurePS}a). Thus,
$A_1:=A_0^*\setminus B_1$.

\item
$(\tilde{H}^1_1,\tilde{H}^1_2):=(0,-\mathbf{1}_{[K_2,b]}(S_1))$ is
an arbitrage aggregator on $A_1$: the price process does not
increase and $(\tilde{H}^1_1,\tilde{H}^1_2)$ yields a positive
gain on $B_2:=\{K_2\leq S_1\leq b\}\setminus \{\{K_2\}\times
\{K_2\}\}$ (see Figure \ref{FigurePS}b). Thus, $A_1^*=A_1\setminus
B_2$.

\item The null process $H^2$ and the vector
$\alpha^2:=(-1,0)\in\R^2$ satisfies
\begin{equation*}
(H^2\circ S)_T+\alpha^2\cdot \Phi\geq 0\text{ on }A_1^*.
\end{equation*}
A positive gain is obtained on $[0,b]\times[0,K_2]$. Thus
$A_2:=(b,4)\times [0,4]\cup \{\{4\}\times \{4\}\}$.

\end{enumerate}
Obviously there are no more semi-static 1p-arbitrage opportunities and $A_2=A_2^*$,
$\beta=2$. We set $\tilde{H}^2\equiv 0$ and the partition scheme is successful with arbitrage
aggregators $\tilde{H}^{0},\tilde{H}^{1},\tilde{H}^{2}$, and
semi-static strategies $(H^j,\alpha^{j})$, for $j=1,2$, as above.
\begin{figure}[h]
\centering
\begin{minipage}{0.45\textwidth}
\begin{tikzpicture}[scale=0.8]
\footnotesize
  \draw[->] (0,-1) -- (5,-1) node[right] {$t$};

  \foreach \x in {0,1,2}
    \draw (\x*2 ,-1) -- (\x*2 ,-1) node[below] {$\x$};

    \draw [fill] (0,2) circle [radius=0.05]node[left] {$2$};
    \draw [|-|](2,0) -- (2,4);
    \draw[|-|] (4,1) -- (4,4);
    \draw (2,4.34) node[above]{ $S_1$};
    \draw (4,4.28) node[above]{$S_2\mid \{S_1=s\}$};
    \draw[|-] (4,0) -- (4,1)node[left] {$K_2$};
    \draw [fill] (2,5/4) circle [radius=0.05] node[left] {$s$};
    \draw[-, dotted](2,5/4)--(4,4);
    \draw[-, dotted](2,5/4)--(4,0);
    \draw[-, dotted](0,2)--(2,0);
    \draw[-, dotted](0,2)--(2,3/2);
    \draw[decorate, decoration={brace,amplitude=8pt}]
    (4.3,4)--(4.3,1)node[midway,right,xshift=0.4cm]{$B_1$};
   \draw (2,-2) node[below]{$a)$};

\end{tikzpicture}

\end{minipage}
\begin{minipage}{0.45\textwidth}
\begin{tikzpicture}[scale=0.8]
\footnotesize
  \draw[->] (0,-0.9) -- (5,-0.9) node[right] {$t$};

  \foreach \x in {0,1,2}
    \draw (\x*2 ,-0.9) -- (\x*2 ,-0.9) node[below] {$\x$};

    \draw [fill] (0,2) circle [radius=0.05]node[left] {$2$};
    \draw [|-|](2,0) -- (2,4);
    \draw (2,4.34) node[above]{$S_1$};
    \draw (4.3,4.27) node[above]{$S_2\mid \{S_1=s, A_1\}$};
    \draw[|-|] (4,0) -- (4,1)node[left] {$K_2$};
    \draw [fill] (2,5/4) circle [radius=0.05] node[left] {$s$};
    \draw[-, dotted](2,5/4)--(4,1);
    \draw[-, dotted](2,5/4)--(4,0);
    \draw[-, dotted](0,2)--(2,0);
    \draw[-, dotted](0,2)--(2,3/2);
    \draw[decorate, decoration={brace,amplitude=3pt}]
    (4.3,1)--(4.3,0)node[midway,right,xshift=0.1cm]{$B_2$};
   \draw (2,-1.9) node[below]{$b)$};

\end{tikzpicture}
\end{minipage}
\caption{Some steps of the pathspace partition scheme with $b=1.5$, $K_2=1$. On the left, the strategy $(H_1,\alpha^1)$, defined on $A_0^*$, has positive gain on $B_1$. As a consequence the pathspace reduces to $A_1=A_0^*\setminus B_1$. On the right, for $S_1\geq K_2$ the arbitrage aggregator $\tilde{H}_1$, defined on $A_1$, has positive gain on $B_2$.}\label{FigurePS}
\end{figure}
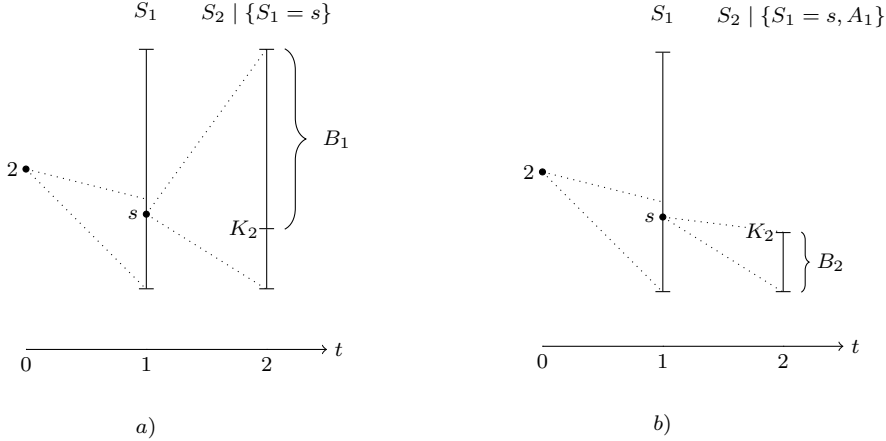

\begin{remark}\label{noWayIfNoDominate} The previous example also shows that $\pi_{\Omega^{\ast }_{\Phi}, \Phi}(g)=\pi_{\Omega, \Phi}(g)$
is a rather exceptional case if we do not assume the existence of
an option with dominating payoff as in Theorem \ref{thm: AB
theorem 2}. Consider indeed the market of example \ref{ex:
partition} with $b:=4$, $0<K_2<2$, which has the same features as
the example on page 5 in \citet{DH07}. Take $g\equiv 1$ and note
that $\Omega$ is compact and $S$, $\Phi$ and $g$ are continuous
functions on $\Omega$. From the above discussion we see easily
that $\Omega^{\ast }_{\Phi}=\emptyset$ and $\pi_{\Omega^{\ast
}_{\Phi},\Phi}=-\infty$. Nevertheless, by considering the
pathspace partition scheme above, we see that while we can devise
an arbitrage strategy on $B_1 = [0,4)\times (K_2,4]$ its payoff is
not bounded below by a positive constant and in fact we see that
$\pi_{\Omega, \Phi}(g)=1$.
\end{remark}

\begin{remark}\label{remark:10}
Note that if a partition scheme is successful then there are no
One-Point Arbitrages on $A^*_\beta$. When $\beta<k$ this follows from $(iv)$ in Definition \ref%
{def:pps}. In the case $\beta=k$ suppose there is a One-Point Arbitrage $(\alpha,H)\in\mathcal{A}_{\Phi}(\prFF)$ so that, in particular, $(H \circ S)_T + \alpha \cdot \Phi \ge 0$ on $A^*_{\beta}$. Since the vectors $\alpha^i$ form a basis of $\R^k$ we get, for some $\lambda_i\in\R$,
$$(H \circ S)_T + \alpha \cdot \Phi=\sum_{i=1}^k\lambda_i\left[ (H^i \circ S)_T + \alpha^i \cdot \Phi\right]+(\hat{H} \circ S)_T$$ where $\hat{H}:=H-\sum_{i=1}^k\lambda_iH^i$. Since, by construction $(H^i \circ S)_T + \alpha^i \cdot \Phi=0$ on $A^*_{\beta}$ for any $i=1,\ldots,\beta$, we obtain that $\hat{H}$ is a One-Point Arbitrage with $\Phi=0$ on $A^*_{\beta}$. From Lemma \ref{NoOnePointNoOptions} we have a contradiction.
\end{remark}

\begin{remark}
\label{rmkNo1pArb} As we shall see, Lemma \ref{A_beta} implies relative uniqueness of $\mathcal{R%
}(\alpha^{\star},H^{\star})$ in the sense that either every $\mathcal{R}%
(\alpha^{\star},H^{\star})$ is not successful or all $\mathcal{R}%
(\alpha^{\star},H^{\star})$ are successful and then $A^*_{\beta} =
\Omega^*_{\Phi}$.
\end{remark}

\begin{definition}
Given a pathspace partition scheme we define the Arbitrage
Aggregator as
\begin{equation}\label{Aggregator:options}
(\alpha^{\star},H^{\star})= \left(\sum_{i=1}^{\beta} \alpha^i\mathbf{1}%
_{A^*_i},\sum_{i=1}^{\beta}H^i\mathbf{1}_{A^*_i}+\sum_{i=1}^{\beta}\tilde{H}%
^i\mathbf{1}_{A_i\setminus A^*_i}\right),
\end{equation}
with $(\alpha^{\star},H^{\star})=(0,\tilde H^0\mathbf{1}%
_{\Omega\setminus \Omega^*})$ if $\beta=0$.
\end{definition}

To make the above arbitrage aggregator predictable we need to
enlarge the
filtration. We therefore introduce the arbitrage aggregating filtration $%
\widetilde{\mathbb{F}}$ given by
\begin{equation}  \label{eq:Ftilde_options}
\begin{split}
\widetilde{\mathcal{F} }_t = & \mathbb{F}^{S,Y}_t\vee
\{A_0,
A^*_0, \ldots,A_{\beta}, A^*_{\beta}\}\vee \sigma(\tilde{H}^0_1,\ldots,%
\tilde{H}^{\beta}_1, \ldots, \tilde{H}^0_{t+1}, \ldots, \tilde
H^{\beta}_{t+1}), \\
\widetilde{\mathcal{F} }_T = & \mathbb{F}^{S,Y}_T\vee
\{A_0,
A^*_0, \ldots,A_{\beta}, A^*_{\beta}\}\vee \sigma(\tilde{H}^0_1,\ldots,%
\tilde{H}^{\beta}_1, \ldots, \tilde{H}^0_{T}, \ldots, \tilde
H^{\beta}_{T}).
\end{split}%
\end{equation}

It will follow, as a consequence of Lemma \ref{polars2}, that $\widetilde{\mathcal{F}}_{t}\subseteq \mathcal{F}^M_{T}$ for any $t=0,\ldots, T$ and in particular, as observed before,
any $\mathbb{Q}  \in \mathcal{M}_{\Omega,\Phi}(\mathbb{F}^S)$
extends uniquely to a measure in $\mathcal{M}_{\Omega,\Phi}(\widetilde{\mathbb{F}})$.

\section{Proofs}\label{sec:proofs}

We first describe the logical flow of our proofs and we point out
that we need to show the results for $\Omega\in\pr$ (and not only analytic). In particular, showing $\Omega^*_\Phi\in \pr$ is
involved. First, Theorem \ref{GFTAP} and then Theorem \ref{thm:
duality_options} are established when $\Phi=0$. Then, we show
Theorem \ref{thm: duality_options} for $\Omega\in\pr$, \emph{under the further
assumption} that $\Omega^*_{\Phi_i}\in \pr$ for all $1\leq i\leq
k$, where $\Phi_i=\{\phi_1,\ldots, \phi_i\}$. Note that in the
case with no statically traded options ($\Phi=0$) the property
$\Omega^*_\Phi=\Omega^*\in \pr$ follows from
the construction and is shown in Lemma \ref{LemNOpolar}. This
allows us to prove Proposition \ref{1p-Arb} for which we use
Theorem \ref{thm: duality_options} only when $\Omega_{\Phi_i
}^{\ast }=\Omega$, which belongs to $\pr$ by
assumption. Proposition \ref{1p-Arb} in turn allows us to
establish Lemma \ref{A_beta} which implies that in all cases
$\Omega^*_{\Phi_i}\in \pr$. This then
completes the proofs of Theorem \ref{GFTAP} and Theorem \ref{thm:
duality_options} in the general setting.

\subsection{Proof of the FTAP and pricing hedging duality when no options are statically traded}
$\,$

\begin{prova}[Proof of Theorem \ref{GFTAP}, when no options are statically
traded] In this case we consider $\Omega\in\pr$, the technical filtration as
described in Remark \ref{technical:0options} and the Arbitrage
Aggregator $H^*$ defined by \eqref{universal}.  We prove that
\begin{equation*}
\exists \text{ Strong Arbitrage on } \Omega \text{ in }
\mathcal{H}(\widetilde{\mathbb{F}})\Leftrightarrow
\mathcal{M}^f_{\Omega}= \emptyset.
\end{equation*}%
Notice that if $H\in \mathcal{H}(\widetilde{\mathbb{F}})$
satisfies $(H\circ
S)_T(\omega)> 0\ \forall \omega \in \Omega $ then, if there exists $Q\in%
\mathcal{M}^f_{\Omega}$ we would get $0<\mathbb{E}_{Q}[(H\circ
S)_T]=0$ which is a contradiction. For the opposite implication,
let $H^*$ be the
Arbitrage Aggregator from \eqref{universal} and note that $%
(H^*\circ S)_T(\omega)\geq 0\ \forall \omega \in \Omega $ and
$\{\omega\mid (H^*\circ S)_T(\omega)>0\}=(\Omega^{\ast })^{c}.$ If
$\mathcal{M}^f_{\Omega}=\emptyset $ then, by Lemma \ref{LemNOpolar},
$(\Omega^{\ast })^{c}=\Omega $ and $H^*$ is therefore a Strong
Arbitrage on $\Omega$  in $\mathcal{H}(\widetilde{\mathbb{F}})$.
The last assertion, namely $\Omega^*=\{\omega\in\Omega\mid
(H^*\circ S)_T(\omega)=0 \}$, follows straightforwardly from the
definition of $H^*$.
\end{prova}

\begin{proposition}[Superhedging on $\Omega \subseteq X$ without options]
\label{thm: duality_no_option} Let $\Omega \in
\pr$. We have that for any $g\in
\mathcal{L}(X,\mathcal{F}^{\mathcal{A}};\R)$
\begin{equation}
\pi _{\Omega ^{\ast }}(g)=\sup_{Q\in \mathcal{M}_{\Omega}^{f}}E_{Q}[g],
\label{super:0opt}
\end{equation}%
with $\pi _{\Omega ^{\ast }}(g)=\inf \left\{ x\in \mathbb{R}\mid
\exists
H\in \mathcal{H}(\prFF)\text{ such that }%
x+(H\circ S)_{T}\geq g\ \text{ on } \Omega ^{\ast }\right\} $. In
particular, the left hand side of \eqref{super:0opt} is attained
by some strategy $H\in \mathcal{H}(\prFF)$.
\end{proposition}

\begin{prova}
Note that by its definition in \eqref{eq:OmegaPhi_def},
$\Omega^*=\emptyset$ if and only if $\mathcal{M}_{\Omega}^{f}=
\emptyset $ and in this case both sides in \eqref{super:0opt} are
equal to $-\infty$. We assume now that
$\Omega^*\neq \emptyset$ and recall from Lemma \ref{LemNOpolar} we have $\Omega^*\in%
\pr$. By definition, there exists $n\in\N$ such that $\Omega^*\in\Sigma^1_n$. The second part of the statement
follows with the same procedure proposed in \citet{BFM16b} proof of
Theorem 1.1. The reason can be easily understood recalling the
following construction, which appears in Step 1 of the proof. For
any $\ell\in\N$, $D\in \prk$, $1\leq t\leq T$, $G\in
\mathcal{L}(X ,\prk),$ we define the
multifunction
\begin{equation*}
\psi _{t,G,D}:\omega \mapsto \left\{ \left[ \Delta S_{t}(\widetilde{\omega }%
);1;G(\widetilde{\omega })\right] \mathbf{1}_{D}(\widetilde{\omega
})\mid
\widetilde{\omega }\in \Sigma _{t-1}^{\omega }\right\} \subseteq \mathbb{R}%
^{d+2}
\end{equation*}%
where $\left[ \Delta S_{t};1;G\right] \mathbf{1}_{D}=\left[ \Delta S_{t}^{1}%
\mathbf{1}_{D},\ldots ,\Delta S_{t}^{d}\mathbf{1}_{D},\mathbf{1}_{D},G%
\mathbf{1}_{D}\right] $ and $\Sigma_t^{\omega}$ is given as in
\eqref{level:set}. We show that $\psi _{t,G,D}$ is an
$\prkp_{t-1}$-measurable multifunction.
Let $O\subseteq
\mathbb{R}^{d}\times \mathbb{R}^{2}$ be an open set and observe that
\begin{equation*}
\{\omega \in X \mid \psi _{t,G,D}(\omega )\cap O\neq \emptyset
\}=(S_{0:t-1},Y_{0:t-1})^{-1}\left( (S_{0:t-1},Y_{0:t-1})\left(
B\right) \right),
\end{equation*}%
where $B=(\left[ \Delta S_{t};1;G\right] \mathbf{1}_{D})^{-1}(O)$.
First $\left[ \Delta S_{t},1,G\right] \mathbf{1}_{D}$ is an $\prk$-measurable random vector then $B\in \prk$, the sigma-algebra generated by the $\ell$-projective sets of $X$. Second $(S_{u},Y_u)$ is a Borel measurable function for any $0\leq
u\leq t-1$ so that we have, as a consequence of Lemma
\ref{Fsigma}, that $(S_{0:t-1},Y_{0:t-1})(B)$ belongs to the
sigma-algebra generated by the ($\ell+1$)-projective sets in
$Mat((d+\tilde{d})\times t;\mathbb{R})$ (the space of
$(d+\tilde{d})\times
t$ matrices with real entries) endowed with its Borel sigma-algebra. Applying again Lemma %
\ref{Fsigma} we deduce that $(S_{0:t-1},Y_{0:t-1})^{-1}\left(
(S_{0:t-1},Y_{0:t-1})\left(
B\right) \right)\in \prkp_{t-1}$ and hence the desired measurability for $\psi
_{t,G,D}$. \newline The remaining of Step 1, Step 2, Step 3, Step
4 and Step 5 follows replicating the argument in \citet{BFM16b}.
\end{prova}

\subsection{Proof of the FTAP and pricing hedging duality with statically traded options}
\label{sec:proof_FTAP}
We first extend the results from Lemma \ref{LemNOpolar} to the present case of non-trivial $\Phi$.
\begin{lemma}\label{polars2}
Let $\Omega$ be analytic. For any $Q\in\mathcal{M}_{\Omega, \Phi}$ we have $Q(\Omega^{\ast}_{\Phi})=1$.
In particular, $\mathcal{M}_{\Omega, \Phi}\neq \emptyset$ if and only if $\mathcal{M}^{f}_{\Omega, \Phi}\neq \emptyset$.
\end{lemma}
\begin{prova}
%
Recall that $\Omega$ analytic implies that $\Omega^\ast_{\phi}$ is analytic, from Remark 5.6 in \citet{BFM16b}. Let $\tilde{Q}\in\mathcal{M}_{\Omega, \Phi}$ and consider the extended market $(S,\tilde{S})$ with $\tilde{S}^j_t$ equal to a Borel-measurable version of $E_{\tilde Q}[\phi_j\mid\F^S_t]$ for any $j=1,\ldots,k$ and $t\in\T$ (see Lemma 7.27 in \citet{BS78}). In particular $\tilde{Q}\in\tilde{\mathcal{M}}_{\Omega}$, the set of martingale measure for $(S,\tilde{S})$ which are concentrated on $\Omega$. From Corollary \ref{polars} and Lemma \ref{LemNOpolar} we deduce that $\tilde{Q}(\tilde{\Omega}^\ast)=1$. Since, obviously, $\tilde{\mathcal{M}}^f_{\Omega}\subseteq \mathcal{M}^f_{\Omega, \Phi}$ we also have $\tilde{\Omega}^\ast\subseteq \Omega^{\ast}_{\Phi}$. Since the former has full probability  the claim follows.\\
\end{prova}

\begin{prova}[Proof of Theorem \ref{thm: duality_options} under the assumption $\Omega^*_{\Phi_n}\in \pr$ for all
$n\leq k$] Let $\Omega\in\pr$. Similarly to the proof of Proposition \ref{thm:
duality_no_option} we note
that the statement is clear when $\mathcal{M}_{\Omega ,\Phi
}^{f}=\emptyset $ so we may assume the contrary.

The equality between the suprema over $\MM_{\Omega,\Phi}$ and over
$\MM^{f}_{\Omega,\Phi}$ may be deduced following the same arguments
as in the proof of Theorem 1.1, Step 2, in \cite{BFM16b}.
It follows that for any $\mathcal{F}^{\mathcal{A}}$-measurable $g$
$$\sup_{Q\in\mathcal{M}_{\Omega ,\Phi }^{f}}E_{Q}[g]\leq \sup_{Q\in\mathcal{M}_{\Omega ,\Phi }}E_{Q}[g] = \sup_{Q\in\mathcal{M}_{\Omega ,\Phi }}\tilde{E}_{Q}[g]\leq \pi _{\Omega _{\Phi }^{\ast },\Phi }(g)
$$
and it remains to show the equality between the first and the last term above, i.e.\ the first equality in \eqref{eq:PHduality_options}.

Recall that $\Phi_n = (\phi_1, \ldots, \phi_{n})$, $1\leq n\leq k$ with $%
\Phi_k = \Phi$. We prove the statement by induction on the number
of static options used for superhedging. For this we consider the
superhedging problem with additional options $\Phi_n$ on
$\Omega^*_{\Phi}$ and denote its superhedging cost by
$\pi_{\Omega^*_{\Phi}, \Phi_n}(g)$ which is defined as in
\eqref{eq:sh_options} but with $\Phi_n$ replacing $\Phi$. \\
\emph{Assume that $\Omega^*_{\Phi_n}\in \pr$ for all $n\leq k$.}
The case $n=0$ corresponds to the super-hedging problem on
$\Omega^*$ when only dynamic trading is possible. Since by
assumption $\Omega^*_{\Phi}\in \pr$, the
pricing--hedging duality and the attainment of the infimum follow
from Proposition \ref{thm: duality_no_option}. Now assume that for
some $n<k$, for any $\F^{\mathcal{A}}$-measurable $g$, we have the
following pricing--hedging duality
\begin{equation}
\pi_{\Omega^*_{\Phi}, \Phi_n}(g) = \sup_{Q\in \mathcal{M}^f_{\Omega^*_{%
\Phi},\Phi_n}}E_{Q}[g]  \label{eq: duality for n}
\end{equation}
We show that the same statement holds for $n+1$. Note that the
attainment property is always satisfied. Indeed  using the
notation of \citet{BN13}, we have
$NA(\MM^{f}_{\Omega^*_{\Phi},\Phi_{n}})$. As a consequence of
Theorem 2.3 in \citet{BN13}, which holds also in the setup of this paper, the infimum is attained  whenever is finite.\\

The proof proceeds in three steps.
\smallskip\newline
\textbf{Step 1. } First observe that if $\phi_{n+1}$ is replicable on $%
\Omega^*_{\Phi}$ by semi-static portfolios with the static hedging
part restricted to $\Phi_n$, i.e. $x+h\cdot\Phi_n(\omega) +
(H\circ S)_T(\omega) = \phi_{n+1}(\omega)$, for any $\omega\in
\Omega^*_{\Phi}$, then necessarily $x=0$ (otherwise
$\mathcal{M}^f_{\Omega,\Phi}=\emptyset$). Moreover since any such
portfolio has zero expectation
under measures in $\mathcal{M}^f_{\Omega^*_{\Phi},\Phi_n}$ we have that $%
E_{Q}[\phi_{n+1}] = 0$ $\forall Q\in
\mathcal{M}^f_{\Omega^*_{\Phi},\Phi_n}$. In particular
$\mathcal{M}^f_{\Omega^*_{\Phi},\Phi_n}=
\mathcal{M}^f_{\Omega^*_{\Phi},\Phi_{n+1}}$ and \eqref{eq: duality
for n}
holds for $n+1$. \\

\textbf{Step 2. } We now look at the more interesting case, that is $%
\phi_{n+1}$ is not replicable. In this case, we show that:
\begin{align}
\begin{split}  \label{pos_neg}
\sup_{Q\in \mathcal{M}^f_{\Omega^*_{\Phi},\Phi_n}}E_{Q}[\phi_{n+1}] >0\quad%
\text{and}\quad\inf_{Q\in \mathcal{M}^f_{\Omega^*_{\Phi},\Phi_n}}E_{Q}[%
\phi_{n+1}] <0.
\end{split}%
\end{align}
Inequalities $\ge$ and $\le$ are obvious from the assumption
$\mathcal{M}^f_{\Omega^*_{\Phi},\Phi}\neq\emptyset$. From the
inductive hypothesis we only need to show that
$\pi_{\Omega^*_{\Phi}, \Phi_n}(\phi_{n+1})$ is always strictly
positive (analogous argument applies to
$\pi_{\Omega^*_{\Phi},\Phi_n}(-\phi_{n+1})$). Suppose, by
contradiction, $\pi_{\Omega^*_{\Phi}, \Phi_n}(\phi_{n+1})=0$.
Since the infimum is attained, there exists some $(\alpha,H)\in
\R^{n}\times\mathcal{H}(\prFF) $ such that
\begin{equation*}
\alpha\cdot\Phi_n(\omega) + (H\circ S)_T(\omega) \ge
\phi_{n+1}(\omega) \quad \forall \omega \in \Omega^*_{\Phi}.
\end{equation*}
Since $\phi_{n+1}$ is not replicable the above inequality is
strict for some $\tilde{\omega}\in \Omega^*_{\Phi}
$. Then, by taking expectation under $\tilde{Q}\in \mathcal{M}%
^f_{\Omega^*_{\Phi},\Phi}$ such that
$\tilde{Q}(\{\tilde{\omega}\})>0$, we obtain
\begin{equation}  \label{eq:proof_sh_contr1}
0=E_{\tilde{Q}}[\alpha\cdot\Phi_n+ (H\circ S)_T] >
E_{\tilde{Q}}[\phi_{n+1}] = 0.
\end{equation}
which is clearly a contradiction.\\

\textbf{Step 3.} Given \eqref{pos_neg}, we now show that
\eqref{eq: duality for n} holds for $n+1$, also in the case that
$\phi_{n+1}$ is not replicable. We first use a variational
argument to deduce the following equalities:
\begin{align}
\pi_{\Omega^*_{\Phi}, \Phi_{n+1}}(g) =&\; \inf_{l\in \R}\pi_{\Omega^*_{%
\Phi}, \Phi_{n}}(g - l\phi_{n+1})  \label{eq: variational equality 1} \\
=&\; \inf_{l\in \R}\sup_{Q\in \mathcal{M}^f_{\Omega^*_{\Phi},%
\Phi_n}}E_{Q}[g- l\phi_{n+1}]  \notag \\
=&\; \inf_{N}\inf_{|l|\le N}\sup_{Q\in \mathcal{M}^f_{\Omega^*_{\Phi},%
\Phi_n}}E_{Q}[g- l\phi_{n+1}]  \notag \\
=&\; \inf_{N}\sup_{Q\in
\mathcal{M}^f_{\Omega^*_{\Phi},\Phi_n}}\inf_{|l|\le
N}E_{Q}[g- l\phi_{n+1}],  \notag \\
=&\; \inf_{N}\sup_{Q\in \mathcal{M}
^f_{\Omega^*_{\Phi},\Phi_n}}\left(E_{Q}[g]-
N|E_{Q}[\phi_{n+1}]|\right) \notag
\end{align}
The first equality follows by definition, the second from the
inductive
hypothesis, the fourth is obtained with an application of min--max theorem (see Corollary 2 in \citet{T72}) and the last one follows from an easy calculation.\\

We also observe that there exist $Q_{\sup}\in \mathcal{M}^f_{\Omega^*_{\Phi},\Phi_n}$ and $%
Q_{\inf}\in \mathcal{M}^f_{\Omega^*_{\Phi},\Phi_n}$ such that
\begin{equation*}
E_{Q_{\sup}}[\phi_{n+1}] \ge \frac{1}{2}\left(
\pi_{\Omega^*_{\Phi}, \Phi_n}(\phi_{n+1})\wedge 1\right) \quad
\text{ and } \quad E_{Q_{\inf}}[\phi_{n+1}] \le -\frac{1}{2}\left(
\pi_{\Omega^*_{\Phi}, \Phi_n}(-\phi_{n+1})\wedge 1\right).
\end{equation*}
From \eqref{pos_neg} and the inductive hypothesis
$E_{Q_{\inf}}[\phi_{n+1}]< 0<E_{Q_{\sup}}[\phi_{n+1}]$. We will
later use $Q_{\inf}$ and $Q_{\sup}$ for calibrating measures in
$\mathcal{M}^f_{\Omega^*_{\Phi},\Phi_n}$ to the additional option
$\phi_{n+1}$. Namely, for
$Q\in\mathcal{M}^f_{\Omega^*_{\Phi},\Phi_n}$ we might set
$\tilde{Q} = Q_{\inf}$ if $E_{Q}[\phi_{n+1}]\ge 0$,
and $Q_{\sup}$ otherwise, to find $%
\lambda\in [0,1]$ such that
\begin{equation*}
\hat{Q} = \lambda Q +(1-\lambda)\tilde{Q}\in \mathcal{M}^f_{\Omega^*_{\Phi}, \Phi_{n+1}}.
\end{equation*}

We can now distinguish two cases:

\textbf{Case 1. }Suppose first there exists a sequence
$\{Q_m\}\subseteq
\mathcal{M}^f_{\Omega^*_{\Phi},\Phi_{n}}\setminus
\mathcal{M}^f_{\Omega^*_{\Phi},\Phi_{n+1}}$ such that
\begin{equation}\label{supInfinity}
\lim_{m\rightarrow\infty}\dfrac{E_{Q_m}[g]}{|E_{Q_m}[\phi_{n+1}]|}=
+\infty\quad \text{ and } \quad \lim_{m\rightarrow\infty}
E_{Q_m}[g]=+\infty.
\end{equation}

Given $\{Q_m\}$ such that \eqref{supInfinity} is satisfied,
we can construct a sequence of calibrated measures $\{\hat{Q}_m\}\subseteq \mathcal{M%
}^f_{\Omega^*_{\Phi},\Phi_{n+1}}$, as described above, so that
$$E_{\hat{Q}_m}[\phi_{n+1}]=\lambda_m E_{Q_m}[\phi_{n+1}]+(1-\lambda_m) E_{\tilde{Q}_m}[\phi_{n+1}]=0,$$
for some $\{\lambda_m\}\subseteq [0,1]$. We stress that $\tilde
Q_m$ can only be equal to $Q_{\inf}$ or $Q_{\sup}$, which do not
depend on $m$. A simple calculation shows
$$\lambda_m=\frac{E_{\tilde{Q}_m}[\phi_{n+1}]}{E_{\tilde{Q}_m}[\phi_{n+1}]-E_{Q_m}[\phi_{n+1}]}. $$

From
$$E_{\hat{Q} _m}[g]=\lambda_m
(E_{Q_m}[g]-E_{\tilde{Q}_m}[g])+E_{\tilde{Q}_m}[g] $$ we have two cases: either $\lambda_m\rightarrow a >0$ and from $E_{Q_m}[g]\to +\infty$ we deduce $E_{\hat{Q} _m}[g]\to +\infty$; or $\lambda_m\rightarrow 0$  which happens when $|E_{Q_m}[\phi_{n+1}]|\to\infty$. Nevertheless in such a case, from \eqref{supInfinity} we obtain again $E_{\hat{Q} _m}[g]\to +\infty$ as $m\to \infty$. Therefore, $\infty=\sup_{Q\in \mathcal{M}%
^f_{\Omega^*_{\Phi},\Phi_{n+1}}}E_{Q}[g]\leq \pi_{\Omega^*_{\Phi}, \Phi_{n+1}}(g)$ and hence the duality.\\

\textbf{Case 2 }We are only left with the case where
\eqref{supInfinity} is not satisfied. For any $N\in\mathbb{N}$, we
define the decreasing sequence $s_N:=\sup_{Q\in \mathcal{M}
^f_{\Omega^*_{\Phi},\Phi_n}}(E_{Q}[g]- N|E_{Q}[\phi_{n+1}]|)$ and
let $\{Q^m_N\}_{m\in\mathbb{N}}$ a sequence realizing the
supremum. If there exists a subsequence $s_{N_j}$ such that
$|E_{Q_{N_j}^m}[\phi_{n+1}]|=0$ for $m>\bar m(N_j)$, then the
duality follows directly from \eqref{eq: variational equality 1}.
Suppose this is not the case. We claim that we can find a sequence
$\{Q_N\} \subseteq \mathcal{M}^{f}_{\Omega^*_{\Phi},\Phi_n}$ such
that
\begin{equation}\label{claimforSH}
\lim_{N\rightarrow\infty}(E_{Q_N}[g]-
N|E_{Q_N}[\phi_{n+1}]|)=\lim_{N\rightarrow\infty}
s_N\qquad\text{and}\qquad
\lim_{N\rightarrow\infty}|E_{Q_N}[\phi_{n+1}]|=0.
\end{equation}
Let indeed
$c(N):=\limsup_{m\rightarrow\infty}E_{Q_N^m}[g]/|E_{Q_N^m}[\phi_{n+1}]|$.
If $\sup_{N\in\mathbb{N}}c(N)=\infty$, since \eqref{supInfinity}
is not satisfied, there exists $m=m(N)$ such that
$|E_{Q_N^{m(N)}}[\phi_{n+1}]|$ converges to $0$ as $N\rightarrow
\infty$, from which the claim easily follows. Suppose now
$\sup_{N\in\mathbb{N}}c(N)<\infty$. Then (by taking subsequences
if needed),
\begin{eqnarray*}
s_N=\lim_{m\to\infty}(E_{Q_N^m}[g]-
N|E_{Q_N^m}[\phi_{n+1}]|)&\leq&
(c(N)-N)\lim_{m\to\infty}|E_{Q_N^m}[\phi_{n+1}]|,
\end{eqnarray*}
 Note that $a_N:=\lim_{m\to\infty}|E_{Q_N^m}[\phi_{n+1}]|$ satisfies $\lim_{N\rightarrow\infty}Na_N<\infty$ otherwise, from \eqref{eq: variational equality 1}, $\pi_{\Omega^*_{\Phi}, \Phi_{n+1}}(g)=-\infty$ which is not possible as, for any $Q\in \mathcal{M}^f_{\Omega^*_{\Phi},\Phi}$, we have that $ \pi_{\Omega^*_{\Phi}, \Phi_{n+1}}(g) \geq E_Q[g]>-\infty$. In particular, $\lim_{N\rightarrow\infty} a_N=0$ and the claim easily follows.

Given a sequence as in \eqref{claimforSH} we now conclude the
proof. It follows from \eqref{eq: variational equality 1} that
\begin{align*}
\pi_{\Omega^*_{\Phi}, \Phi_{n+1}}(g) =&\; \inf_{N}\sup_{Q\in \mathcal{M}%
^f_{\Omega^*_{\Phi},\Phi_n}}\inf_{|l|\le N}E_{Q}[g- l\phi_{n+1}] \\
=&\; \lim_{N\to\infty}E_{Q_N}[g]- N|E_{Q_N}[\phi_{n+1}]| \\
\le&\; \lim_{N\to\infty}E_{Q_N}[g].
\end{align*}
The calibrating procedure described above yields $%
\lambda_N\in [0,1]$ such that $\hat{Q}_N = \lambda_N Q_{N} +(1-\lambda_N)%
\tilde{Q}_{N}\in \mathcal{M}^f_{\Omega^*_{\Phi}, \Phi_{n+1}}$. Moreover, as $%
|E_{Q_N}[\phi_{n+1}]|\rightarrow 0$ and $\tilde{Q}_N$ can only be
either $Q_{\inf}$ or $Q_{\sup}$, these $\lambda_N$ satisfy
$\lambda_N \to 1$. This implies, $E_{\hat{Q}_N}[g] - E_{Q_N}[g]
\to 0$ as $N\to \infty$ from which it follows
\begin{equation*}
\pi_{\Omega^*_{\Phi}, \Phi_{n+1}}(g) \le \; \lim_{N\to \infty}
E_{Q_N}[g] = \lim_{N\to \infty} E_{\hat{Q}_N}[g] \le \sup_{Q\in
\mathcal{M}^f_{\Omega^*_{\Phi},\Phi_{n+1}}}E_{Q}[g].
\end{equation*}
The converse inequality follows from standard arguments and hence
we obtain $\pi_{\Omega^*_{\Phi}, \Phi_{n+1}}(g) = \sup_{Q\in
\mathcal{M}^f_{\Omega^*_{\Phi},\Phi_{n+1}}}E_{Q}[g]$ as required.
\end{prova}

\bigskip

We now prove Proposition \ref{1p-Arb} for the more general case of $\Omega\in\pr$. We use Theorem
\ref{thm: duality_options} only when
$\Omega^{\ast}_{\Phi}=\Omega$, which belongs to
$\pr$ by assumption.

\bigskip

\begin{prova}[Proof of Proposition \ref{1p-Arb}]
``$\Leftarrow$" is clear since, if a strategy $(\alpha,H)\in \A_\Phi(\prFF)$
satisfies $\alpha\cdot \Phi+(H\circ S)_T\geq 0$ on $\Omega$ then, by definition in
\eqref{eq:OmegaPhi_def}, for any $\omega\in
\Omega^*_{\Phi}=\Omega$, we can take a calibrated martingale measure which
assigns a positive probability to $\omega$, which implies $\alpha\cdot \Phi+(H\circ S)_T= 0$ on $\omega$. Since $\omega$ is arbitrary we obtain the thesis. We prove ``$\Rightarrow$" by iteration on number of
options used for static trading. No One-Point Arbitrage using dynamic
trading and $\Phi$ in particular means that there is no
One-Point Arbitrage using only dynamic trading. From Lemma \ref{NoOnePointNoOptions} we have
$\Omega^*=\Omega$ and hence for
any $\omega\in \Omega$ there exists $Q\in \mathcal{M}^f_{\Omega}$ such that $%
Q(\{\omega\})>0$. \\Note that if, for some $j\leq k$, $\phi_j$ is
replicable on $\Omega^*$ by dynamic trading in $S$ then there exist $n\in\N$ and $(x,H)\in\R\times\mathcal{H}(\prN)$ such that $x+(H\circ S)_T=\phi_j$ on $\Omega^*$. No One-Point Arbitrage implies $x=0$ and hence
$E_Q[\phi_j]=0$ for every $Q\in \mathcal{M}^f_{\Omega}$. With no
loss of generality we assume
that $(\phi_1,\ldots,\phi_{k_1})$ is a vector of non-replicable options on $%
\Omega^*$ with $k_1\leq k$. We now apply Theorem \ref{thm:
duality_options} in the case with $\Phi=0$ to $\phi_1$ and argue
that
\begin{equation*}
m_1:=\min\{\pi_{\Omega^*}( \phi_1),\pi_{\Omega^*}( -\phi_1)\}>0.
\end{equation*}
Indeed, if $m_1<0$ then we would have a Strong Arbitrage and if
$m_1=0$, since the superhedging price is attained, there exists
$H\in \mathcal{H}(\prFF)$ such that, for example, $\phi_1
\leq (H\circ S)_T$ on $\Omega$. In order to avoid One-Point
Arbitrage we have to have $\phi_1=(H\circ S)_T$ on $\Omega$ which
is a contradiction since $\phi_1$ is not replicable. This shows
that $m_1>0$ which in turn implies there exist $Q_1,Q_2\in
\mathcal{M}^f_{\Omega}$ such that $E_{Q_1}[\phi_1]>0$ and
$E_{Q_2}[\phi_1]<0$. Then, for any $Q\in
\mathcal{M}^f_{\Omega}$, there exist $\alpha,\beta,\gamma\in [0,1]$, $%
\alpha+\beta+\gamma=1$ and $E_{\alpha Q_1+\beta Q_2+\gamma
Q}[\phi_1]=0$.
Thus, for any $\omega\in \Omega^*$ there exists $Q\in \mathcal{M}%
^f_{\Omega,\phi_1}$ such that $Q(\{\omega\})>0$. In particular, $%
\Omega^*_{\phi_1}=\Omega$ and we may apply Theorem \ref{thm:
duality_options} with $\Omega$ and $\Phi=\{\phi_1\}$ (Indeed
$\Omega\in \pr$ and we can therefore apply the
version of Theorem \ref{thm: duality_options} proved in this
section). Define now
\begin{equation*}
m_{1,j}:=\min\{\pi_{\Omega^*,\phi_1,}(
\phi_j),\pi_{\Omega^*,\phi_1}( -\phi_j)\} \quad\forall
j=2,\ldots,k_1 .
\end{equation*}
By absence of Strong Arbitrage we necessarily have $%
m_{1,j}\geq 0$ for every $j=2,\ldots,k_1$. Let $j\in
I_2=\{j=2,\ldots,k_1\mid m_{1,j}=0\}$, by No One-Point Arbitrage, we have
perfect replication of $\phi_j$ using semistatic strategies with
$\phi_1$ on $\Omega$
and in consequence for any $Q\in \mathcal{M}^f_{\Omega,\phi_1}$ we have $%
E_Q[\phi_j]=0$ for all $j\in I_2$. We may discard these options
and, up to re-numbering, assume that $(\phi_2,\ldots,\phi_{k_2})$
is a vector of the
remaining options, non-replicable on $\Omega$ with semistatic trading in $%
\phi_1$, with $k_2\leq k_1$. If $k_2\geq 2$, $m_{1,2}>0$ by
Theorem \ref{thm: duality_options} and absence of One-Point
Arbitrage using arguments as above.
Hence, there exist $Q_1,Q_2\in \mathcal{M}^f_{\Omega,\phi_1}$ such that $%
E_{Q_1}[\phi_2]>0$ and $E_{Q_2}[\phi_2]<0$. As above, this implies that $%
\Omega^*_{\{\phi_1,\phi_2\}}=\Omega^*_{\phi_1}=\Omega$. We can
iterate the above
arguments and the procedure ends after at most $k$ steps showing $%
\Omega^*_\Phi=\Omega$ as required.
\end{prova}

\bigskip

The following Lemma shows that the outcome of a successful
partition scheme is the set $\Omega^{\ast}_{\Phi}$.

\begin{lemma}
\label{A_beta} Recall the definition of $\Omega^*_\Phi$ in %
\eqref{eq:OmegaPhi_def}. For any $\mathcal{R}(\alpha^{\star},H^{\star})$%
, $A^*_{i} = \Omega^*_{\{\alpha^j\cdot \Phi\,:\, j\le i\}}$ for
any $i\le \beta$. Moreover, if
$\mathcal{R}(\alpha^{\star},H^{\star})$ is successful, then
$A^*_{\beta} = \Omega^*_{\Phi}$.
\end{lemma}

\begin{prova}
If $\Omega^* = \emptyset$, then the claim holds trivial. We now assume $%
\Omega^* \neq \emptyset$, fix a partition scheme $\mathcal{R}%
(\alpha^{\star},H^{\star})$ and prove the claim by induction on
$i$. For simplicity of notation, let $\Omega^*_i:=
\Omega^*_{\{\alpha^j\cdot \Phi\,:\, j\le i\}}$ with
$\Omega_0=\Omega$. By definition of $A_0$ we have
$A_0^*=\Omega^*=\Omega^*_0$. Suppose now $A^*_{i-1} =
\Omega^*_{i-1}$ for some $i\leq\beta$. Then, by
definition of $\Omega_i$ we have, $\Omega^*_i\subseteq \Omega^*_{i-1}=A^*_{i-1}$. Further, since $%
(H_i \circ S)_T + \alpha^i \cdot \Phi \ge 0$ on $A^*_{i-1}$ with
strict inequality on $ A_{i-1}^*\setminus A_i$, it follows that
$\Omega^*_i\subseteq A_i$. Finally, from $\mathcal{M}^f_{\Omega^*_i,
\{\alpha^j\cdot \Phi\,:\, j\le i\}}\subseteq\mathcal{M}^f_{A_i,
\{\alpha^j\cdot \Phi\,:\, j\le i\}}\subseteq
\mathcal{M}^f_{A_i}=\mathcal{M}^f_{A_i^*}$ we also have
$\Omega^*_i\subseteq A_i^*$. For the reverse inclusion consider
$\omega\in A^*_i$. By definition of $A^*_i$ and Lemma
\ref{LemNOpolar}, there exists $Q\in \mathcal{M}^f_{A_i^*}$ with
$Q(\{\omega\})>0$. Since on $A_i^*$, all options $\alpha^j\cdot
\Phi$, $1\leq j \leq i$, are perfectly replicated by the dynamic
strategies $-H^j$, it follows that $Q \in \mathcal{M}^f_{A_i^*,
\{\alpha^j\cdot \Phi\,:\, j\le i\}}$ so that $\omega\in
\Omega^*_i$. \newline Suppose now
$\mathcal{R}(\alpha^{\star},H^{\star})$ is successful. In
the case $\beta = k$, since $\alpha^i$ form a basis of $\R^k$ we have
$\mathcal{M}^f_{A_\beta^*, \Phi}=\mathcal{M}^f_{A_\beta^*, \{\alpha^j\cdot \Phi\,:\, j\le \beta\}}$
and hence $\Omega^*_\Phi=A^*_\beta$ from the above. Suppose $%
\beta<k$ so that the above shows $\Omega^*_{\Phi}\subseteq
\Omega^*_\beta = A^*_{\beta}$. Observe that since $A_\beta\in
\pr$, from Lemma \ref{LemNOpolar}, we have
$A^*_\beta\in \pr$, moreover, by Remark
\ref{remark:10}, there are no One-Point Arbitrages on $A^*_\beta$.
Thus each $\omega\in A^*_\beta$ is weighted by some $Q\in
\mathcal{M}^f_{A_\beta^*, \Phi}\subseteq \mathcal{M}^f_{\Omega,
\Phi}$ by Proposition \ref{1p-Arb} applied to $A^*_\beta$.
Therefore $A^*_\beta\subseteq \Omega^*_{\Phi}$, which concludes
the proof.
\end{prova}

\begin{remark}\label{split} It follows from Lemma \ref{LemNOpolar} that
$A_{i}^* \equiv \Omega^*_{\{\alpha^j\cdot \Phi\,:\, j\le i\}}$
introduced in Lemma \ref{A_beta},
belongs to $\pr$ for any $i\le \beta$. In particular $\Omega^*_\Phi$ in %
\eqref{eq:OmegaPhi_def} is in $\pr$ (see also the discussion after Definition \ref{def:pps}).
\end{remark}

\begin{prova}[Proof of Theorem \ref{GFTAP}] We now prove the pointwise Fundamental Theorem of Asset Pricing,
when semistatic trading strategies in a finite number of options
are allowed. Let $\Omega$ be analytic, $\mathcal{R}%
(\alpha^{\star},H^{\star})$ be a pathspace partition scheme and $%
\widetilde{\mathbb{F}} $ be given by \eqref{eq:Ftilde_options}.
We first show that the following are equivalent:

\begin{enumerate}
\item \label{item_success} $\mathcal{R}(\alpha^{\star},H^{\star})$
is successful.

\item \label{item_Mf} $\mathcal{M}^f_{\Omega, \Phi} \neq
\emptyset$.

\item \label{item_M} $\mathcal{M}_{\Omega, \Phi} \neq \emptyset$.

\item \label{item_NMIA} No Strong Arbitrage with respect to $%
\widetilde{\mathbb{F}}$.
\end{enumerate}
\eqref{item_success} $\Rightarrow$ \eqref{item_Mf} follows from
Remark \ref{rmkNo1pArb} and the definition
of $\Omega^*_\Phi$ in \eqref{eq:OmegaPhi_def}, since $A_{\beta}^*=\Omega_{\Phi}^*$. \eqref{item_Mf} $\Leftrightarrow$ %
\eqref{item_M} follows from Lemma \ref{polars2}. To show \eqref{item_Mf} $\Rightarrow$ %
\eqref{item_NMIA}, observe that under $Q\in
\mathcal{M}^f_{\Omega,\Phi}$ the expectation of any admissible
semi-static trading strategy is zero which excludes the
possibility of existence of a Strong Arbitrage. For the
implication \eqref{item_NMIA} $\Rightarrow$ \eqref{item_success}
note that, for $1\le i\le \beta$, we have $(\tilde{H}^{i-1} \circ
S)_T>0$ on $A_{i-1}\setminus A^*_{i-1}$ from the properties of the
Arbitrage Aggregator (see Remark \ref{strictlyPos}) and $(H^i
\circ S)_T + \alpha^i \cdot \Phi > 0$ on $A_{i-1}^*\setminus A_i$
by construction, so that a positive gain is realized on
$A_{i-1}\setminus A_i$. Finally, from
$\Omega=A_0=(\cup_{i=1}^{\beta} A_{i-1}\setminus A_{i})\cup
A_{\beta}$ and $(\tilde{H}^{\beta} \circ S)_T>0$ on
$A_{\beta}\setminus A^*_{\beta}$, we get
\begin{equation}\label{eq: posArbitrage}
\sum_{i=1}^{\beta}(H^i \circ S)_T + \alpha^i \cdot \Phi +
\sum_{i=0}^{\beta}(\tilde{H}^i\circ S)_T> 0\quad\text{ on }(A^*_{\beta})^C\ ,
\end{equation}and equal to $0$ otherwise.
The hypothesis \eqref{item_NMIA} implies therefore that
$A_{\beta}^*$ is non-empty and hence the pathspace partition
scheme is successful.\\
The existence of the technical
filtration and Arbitrage Aggregator are provided explicitly by \eqref{eq:Ftilde_options} and \eqref{Aggregator:options}. Moreover, from Lemma \ref{A_beta}, $\Omega^*_\Phi=A^*_\beta$. Finally, equation \eqref{def:OmegaPhi} follows from \eqref{eq: posArbitrage}.

\end{prova}

\begin{prova}[Proof of Corollary \ref{SFTAP}]Let $\widetilde\FF$ given by \eqref{eq:Ftilde_options}.
We prove that
\begin{equation*}
\exists \text{ an Arbitrage de la classe }\mathcal{S}\text{ in
}\A_\Phi(\widetilde\FF)\Longleftrightarrow
\mathcal{M}^f_{\Omega,\Phi}=\emptyset \text{ or }
\mathcal{N}^M \text{ contains sets of
}\mathcal{S}.
\end{equation*}

$(\Rightarrow)$: let $(\alpha,H)\in \A_\Phi(\widetilde{\FF})$ be
an Arbitrage de la classe $\mathcal{S}$. By definition, $\alpha\cdot
\Phi+(H\circ S)_T\geq 0$ on $\Omega$ and there exists $A\in\mathcal{S}$ such that $A\subseteq\{\omega\in
\Omega\mid\alpha\cdot \Phi+(H\circ S)_T> 0 \}$.  Note now that for any $Q\in \mathcal{M}^f_{\Omega,\Phi}$ we
have $E_{Q}[\alpha\cdot \Phi+(H\circ S)_T]=0$ which implies
$Q(\{\omega\in \Omega\mid \alpha\cdot \Phi+(H\circ S)_T
> 0 \})=0$. Thus, if $\{\omega\in
\Omega\mid\alpha\cdot \Phi+(H\circ S)_T> 0 \}=\Omega$ then $\mathcal{M}^f_{\Omega,\Phi}=\emptyset$, otherwise, $A\in\mathcal{N}^M\cap \mathcal{S}$.

$(\Leftarrow)$: Consider the
Arbitrage Aggregator $(\alpha^*,H^*)$ as constructed in
\eqref{eq:Ftilde_options} which is predictable with respect to
$\widetilde\FF$ given by \eqref{eq:Ftilde_options}. Let $A \in
\mathcal{N}^M\cap \mathcal{S}$ then, from \eqref{def:OmegaPhi} in Theorem \ref{GFTAP},
$A\subseteq \{\omega\in \Omega\mid \alpha^*\cdot \Phi+(H^*\circ S)_T
> 0 \}$ which implies the thesis.
\end{prova}

\paragraph{End of the Proof of Theorem \ref{thm:
duality_options}.} $\,$ \\As a consequence of Lemma \ref{A_beta},
see also Remark \ref{split}, we obtain: $\Omega^*_{\Phi_n}\in
\pr$ for all $n\leq k$. Therefore the assumption made in the
proof of Theorem \ref{thm: duality_options} at the beginning of
this subsection is always satisfied and the proof is complete.

\subsection{Proof of Theorem \ref{thm: AB theorem 2}}

We recall that the option $\phi_0$ can be only bought at time
$t=0$ and the notations are as follows:
$\widetilde{\mathcal{A}}_{\phi_0}(\prFF):=\{(\alpha,H)\in
\R_+\times \mathcal{H}(\prFF)\}$ and
$\widetilde{\MM}_{\Omega, \phi_0}:=\{Q\in \mathcal{M}_{\Omega} \mid
E_Q[\phi_0]\leq 0\}$.

We first extend the results of Theorem \ref{thm: duality_options} to the case where only $\phi_0$ is available for static trading.
\begin{lemma}\label{Lemma:phi0} Suppose $\widetilde{\MM}^f_{\Omega, \phi_0} \neq \emptyset$, $\pi_{\Omega^*}(\phi_0)>0$ and $\Omega^*_{\phi_{0}}\in\F^\A$. Then, for any  $\mathcal{F}^{\mathcal{A}}$-measurable $g$,
$$\pi_{\Omega^*_{\phi_0} , \phi_0}(g)=\sup_{Q\in \widetilde{\MM}^f_{\Omega, \phi_0}} E_{Q}[g]= \sup_{Q\in \MM^f_{\Omega, \phi_0}} E_{Q}[g].$$
\end{lemma}

\begin{prova}
Note now that the assumption $\pi_{\Omega^*}(\phi_0)>0$
automatically implies $\sup_{Q\in \MM^{f}_{\Omega}}E_{Q}[\phi_{0}]
> 0$. Moreover, by assumption, $\widetilde{\MM}^f_{\Omega,
\phi_0}\neq\emptyset$ from which  $\inf_{Q\in
\MM^{f}_{\Omega}}E_{Q}[\phi_{0}] \leq 0$. \\The idea of the proof
is the same of Theorem \ref{thm: duality_options}. Suppose first
that
\begin{align}
\begin{split}  \label{pos_neg2}
\inf_{Q\in \MM^{f}_{\Omega}}E_{Q}[\phi_{0}] <0 \ .
\end{split}
\end{align}
Then it is easy to see that $\Omega^*_{\phi_0}=\Omega^*$.
We use a variational argument to deduce the following equality:
\begin{equation*}
\pi_{\Omega^*_{\phi_0} , \phi_0}(g) =\pi_{\Omega^* , \phi_0}(g) =\inf_{N}\sup_{Q\in \mathcal{M}
^f_{\Omega}}\left(E_{Q}[g]- N|E_{Q}[\phi_{0}]|\right)
\end{equation*}
obtained with an application of min--max theorem (see Corollary 2
in \citet{T72}). The last Step of the proof of Theorem \ref{thm:
duality_options} is only based on this variational equality and
the analogous of \eqref{pos_neg2} joint with $\sup_{Q\in
\MM^{f}_{\Omega}}E_{Q}[\phi_{0}]
> 0$. By repeating the
same argument we obtain $\pi_{\Omega^*_{\phi_0} , \phi_0}(g)
=\sup_{Q\in \MM^f_{\Omega, \phi_0}} E_{Q}[g]$. Since obviously
\begin{equation}\label{aaa}\pi_{\Omega^*_{\phi_0} , \phi_0}(g)=\sup_{Q\in \MM^f_{\Omega,
\phi_0}} E_{Q}[g]
\leq \sup_{Q\in \widetilde{\MM}^f_{\Omega, \phi_0}} E_{Q}[g]\leq \pi_{\Omega^*_{\phi_0} , \phi_0}(g)
\end{equation} we have the thesis.

Suppose now $\inf_{Q\in \MM^{f}_{\Omega}}E_{Q}[\phi_{0}]=0$. From
Proposition \ref{thm: duality_no_option},
$\pi_{\Omega^*}(-\phi_{0}) =0$ and there exists a strategy
$\bar{H}\in\mathcal{H}(\prFF)$ such that $(\bar
H\circ S)_T\geq -\phi_{0}$ on $\Omega^*$. We claim that the
inequality is actually an equality on $\Omega^*_{\phi_{0}}$ (which
is non-empty by assumption). If indeed for some
$\omega\in \Omega^*_{\phi_{0}}$ the inequality is strict then, any $Q\in \MM^{f}_{\Omega}$ such that $Q(\{\omega\})>0$, satisfies $E_Q[\phi_{0}]>0$, which contradicts $\omega\in \Omega^*_{\phi_{0}}$.
This implies that $\phi_0$ is replicable on $\Omega^*_{\phi_0}$ and thus, $\widetilde{\MM}^f_{\Omega, \phi_0}=\MM^f_{\Omega, \phi_0}=\MM^{f}_{\Omega^*_{\phi_0}}$. In such a case,
\begin{eqnarray*}
\pi_{\Omega^*_{\phi_{0}}, \phi_{0}}(g)\leq \pi_{\Omega^*_{\phi_0}}(g) =\sup_{Q\in \MM^{f}_{\Omega^*_{\phi_0}}}  E_{Q}[g]= \sup_{Q\in \MM^f_{\Omega, \phi_0}}  E_{Q}[g],
\end{eqnarray*}
where, the first equality follows from Proposition \ref{thm: duality_no_option}, since $\Omega^*_{\phi_{0}}\in\F^\A\subseteq\pr$ by assumption. The thesis now follows from standard arguments as above.\\
\end{prova}

\begin{proposition}\label{prop: AB theorem 1}
Assume that $\Omega$ satisfies that there exists an $\omega^*$
such that $S_0(\omega^*) = S_1(\omega^*) =\ldots=S_T(\omega^*)$,
$\Omega =\Omega^*$ and $\pi_{\Omega^*}(\phi_0)>0$. Then the
following are equivalent:
\begin{enumerate}
\item[(1)] There is no Uniformly Strong Arbitrage on $\Omega$ in
$\widetilde{\mathcal{A}}_{\phi_0}(\prFF)$;
\item[(2)] There is no Strong Arbitrage on $\Omega$ in
$\widetilde{\mathcal{A}}_{\phi_0}(\prFF)$;
\item[(3)] $\widetilde{\MM}_{\Omega, \phi_0} \neq \emptyset.$
\item[(4)] $\widetilde{\MM}^f_{\Omega, \phi_0} \neq \emptyset.$
\end{enumerate}
Moreover, when any of these holds,
 for any upper semi-continuous
 $g:\R^{d\times (T+1)}_+\to \R$ such that
 \begin{equation}\label{eq: payoff_dominated}
 \lim_{|x|\to \infty}\frac{g^+(x)}{m(x)} = 0,
 \end{equation}
 where $m(x_0, . . . , x_T ) :=\sum_{t=0}^T g_0(x_t)$,
we have the
following pricing--hedging duality:
\begin{equation} \label{eq: ph-duality one option}
\pi_{\Omega^{\ast }, \phi_0}(g(S))= \sup_{Q\in
\widetilde{\MM}_{\Omega, \phi_0}} E_{Q}[g(S)]= \sup_{Q\in
\MM_{\Omega, \phi_0}} E_{Q}[g(S)].
\end{equation}
\end{proposition}

\begin{remark} We observe that the assumption
$\pi_{\Omega^*}(\phi_0)>0$ is not binding and can be removed. In
fact if $\pi_{\Omega^{\ast }}(\phi_0)\leq 0$, (1) $\Rightarrow$ (3) is obviously
satisfied since $\widetilde{\MM}_{\Omega,
\phi_0}=\MM_{\Omega}\neq\emptyset$. The difference is that the pricing--hedging duality \eqref{eq: ph-duality one option} is (trivially) satisfied only in the first equation.
\end{remark}

\begin{prova}[Proof of Proposition \ref{prop: AB theorem 1}]

(3) $\Rightarrow$ (2) and (2) $\Rightarrow$ (1) are obvious. (4) $\Leftrightarrow$ (3) is an easy consequence of Theorem \ref{GFTAP}. To
show (1) $\Rightarrow$ (4), we suppose there is no Uniformly
Strong Arbitrage on $\Omega$ in
$\widetilde{\mathcal{A}}_{\phi_0}(\prFF)$.

We first show that the interesting case is $\pi_{\Omega^{\ast
}}(\phi_0)> 0$ and $\pi_{\Omega^*}( -\phi_0)= 0$. The other cases follows trivially from Proposition \ref{thm: duality_no_option} and Lemma \ref{Lemma:phi0}:
\begin{itemize}
\item If $\pi_{\Omega^{\ast }}(-\phi_0) < 0$, since the
superhedging price is attained and $\Omega = \Omega^*$, there
exist $H\in \prFF$ and $x<0$ such that
$$  \phi_0(\omega) + (H\circ S)_T(\omega) \ge -x>0, \quad \forall \omega\in \Omega  $$
which is clearly a Uniform Strong Arbitrage on $\Omega$. \item If
$\pi_{\Omega^{\ast }}(-\phi_0) > 0$ and $\pi_{\Omega^{\ast
}}(\phi_0)> 0$, we have that $0$ is in the interior of the price
interval formed by $\inf_{Q\in \mathcal{M}^f_{\Omega}}E_{Q}[
\phi_{0}]$ and $\sup_{Q\in \mathcal{M}^f_{\Omega}}E_{Q}[
\phi_{0}]$. Thus, $\widetilde{\MM}^f_{\Omega,\phi_0}\supseteq
\MM^f_{\Omega,\phi_0}\neq \emptyset$ and, it is straightforward to
see that $\Omega^*_{\phi_0} = \Omega^*$.
\end{itemize}
Note that in all these cases $\Omega^*_{\phi_0}
= \Omega^*=\Omega\in\F^{\A}$ and hence \eqref{eq: ph-duality one option} follows from Lemma \ref{Lemma:phi0}.\\

The remaining case is $\pi_{\Omega^{\ast
}}(\phi_0)> 0$ and $\pi_{\Omega^*}(-\phi_0)= 0$. In this case,
by considering the $\omega^*$ such that $s_0 = S_0(\omega^*) =
S_1(\omega^*) = \ldots = S_T(\omega^*)$, we observe that the super-replication of $-\phi_0$
necessarily requires an initial capital of, at least, $-g_0(s_0)$. From $\pi_{\Omega^*}(-\phi_0)= 0$ we can rule out the
possibility that $g_0(s_0) < 0$. Note now that, by the convexity of $g_0$, for any $l\in {0,\ldots, T-1}$,
$g_0(S_T(\omega)) -
\sum_{i=l+1}^{T}g_0^{\prime}(S_{i-1}(\omega))\Big(S_{i}(\omega) -
S_{i-1}(\omega)\Big) \geq g_0(S_{l}(\omega))$ for any $\omega\in
\Omega$. In particular, when $l = 0$,
\begin{equation}\label{eq: convexity}
 g_0(S_T(\omega)) - \sum_{i=1}^{T}g_0^{\prime}(S_{i-1}(\omega))\Big(S_{i}(\omega) - S_{i-1}(\omega)\Big) \ge g_0(s_0) \quad \forall\omega\in \Omega.
\end{equation}
Denote by $\bar{H}$ the dynamic strategy in \eqref{eq: convexity}.
If $g_0(s_0) > 0$, $(1,\bar{H})$ is a Uniformly Strong Arbitrage
on $\Omega$ and hence, a contradiction to our assumption.  Thus,
$g_0(s_0) = 0$. In this case, it is obvious that the Dirac measure
$\delta_{\omega^*}\in\MMF_{\Omega,\phi_0}\subseteq \widetilde{\MM}^f_{\Omega,\phi_0}$ which is therefore non-empty.\\

 Moreover, since
$\delta_{\omega_*}\in \MMF_{\Omega, \phi_0}$,
\begin{equation*}
\sup_{Q\in \widetilde{\MM}^f_{\Omega, \phi_0}}E_Q[g(S)] \geq\sup_{Q\in \MMF_{\Omega, \phi_0}}E_Q[g(S)]\geq g(s_0, \ldots, s_0).
\end{equation*}

\textbf{Case 1.} Suppose $g$ is bounded from above. We show that it is possible to super-replicate $g$ with any
initial capital larger than $g(s_0, \ldots, s_0)$. To see this recall that, from the strict convexity of $g_0$, the inequality in \eqref{eq: convexity} is strict for any $s\in \R^{d\times (T+1)}_+$ such that
$s$ is not a constant path, i.e., $s_i\neq s_0$ for some $i\in
\{0,\ldots, T\}$. In fact, it is bounded away from $0$ outside any
small ball of $(s_0,\ldots, s_0)$. Hence, due to the upper
semi-continuity and boundedness of $g$, for any $\varepsilon>0$,
there exists a sufficiently large $K$ such that
$$g(s_0, \ldots, s_0) + \varepsilon + K\Big\{g_0(S_T(\omega)) - \sum_{i=1}^{T}g_0^{\prime}(S_{i-1}(\omega))\big(S_{i}(\omega) - S_{i-1}(\omega)\big)\Big\} \ge g(S(\omega)) \quad \forall \omega\in \Omega^*$$
Therefore, $\pi_{\Omega^*, \phi_{0}}(g) \le g(s_0, \ldots, s_0)
\le \sup_{Q\in \MMF_{\Omega, \phi_0}}E_Q[g(S)]\le \sup_{Q\in
\widetilde{\MM}^f_{\Omega, \phi_0}}E_Q[g(S)]$. The converse
inequality is easy and hence we obtain $\pi_{\Omega^*,
\phi_{0}}(g) =\sup_{Q\in \MMF_{\Omega, \phi_0}}E_Q[g(S)]=
\sup_{Q\in \widetilde{\MM}^f_{\Omega,\phi_0}}E_{Q}[g]$ as
required.

\textbf{Case 2.} It remains to argue that the duality still holds
true for any $g$ that is upper semi-continuous and satisfies
\eqref{eq: payoff_dominated}. We first argue that any upper
semi-continuous $g:\R^{d\times (T+1)}_+\to \R$, satisfying
\eqref{eq: payoff_dominated}, can be super-replicated on
$\Omega^*$ by a strategy involving dynamic trading in $S$, static
hedging in $g_0$ and cash. Define a synthetic option with payoff
$\widetilde{m}:\R^{d\times (T+1)}_+ \to \R$ by
\begin{equation}\label{eq: wt m}
\widetilde{m}(x_0, \ldots, x_T) = \sum_{l=0}^{T}\Big\{g_0(x_T) -
\sum_{i=l+1}^{T}g_0^{\prime}(x_{i-1})(x_i - x_{i-1})\Big\}.
\end{equation}
By convexity of $g_0$, we know that
$$ \widetilde{m}(x_0, \ldots, x_T) = \sum_{l=0}^{T}\Big\{g_0(x_T) -
\sum_{i=l+1}^{T}g_0^{\prime}(x_{i-1})(x_i - x_{i-1})\Big\}\ge
\sum_{l=0}^{T}g_0(x_l) = m(x_0, \ldots, x_T).$$ Since we assume
there is no Uniform Strong Arbitrage, it is clear that
$\pi_{\Omega^*, \phi_0}(\widetilde{m}(S)) = 0$.

From \eqref{eq: payoff_dominated} it follows that $g(S) - \widetilde{m}(S)$ is bounded from above. By sublinearity of $\pi_{\Omega^*, \phi_0}(\cdot)$, we have
\begin{align*}
\pi_{\Omega^*, \phi_0}(g) \le&\, \pi_{\Omega^*, \phi_0}(g(S) - \widetilde{m}(S)) + \pi_{\Omega^*, \phi_0}(\widetilde{m}(S))\\
=&\, \sup_{Q\in \MM^f_{\Omega, \phi_0}}E_{Q}[g(S) - \widetilde{m}(S)] + 0\\
=&\, \sup_{Q\in \MM^f_{\Omega, \phi_0}}E_{Q}[g(S)]\leq
\sup_{Q\in \widetilde{\MM}^f_{\Omega, \phi_0}}E_{Q}[g(S)],
\end{align*}
where the first equality follows from the pricing--hedging
duality for claims bounded from above, that we established in Case 1, and the fact
that $\pi_{\Omega^*, \phi_0}(\widetilde{m}(S)) = 0$. Moreover, for $Q\in \MM_{\Omega, \phi_0}$,  $E_Q[\widetilde{m}(S)]=0$ from which the second equality follows.

The converse inequality follows from standard arguments and hence
we have obtained $\pi_{\Omega^*, \phi_{0}}(g) = \sup_{Q\in
\MM^f_{\Omega,\phi_0}}E_{Q}[g]= \sup_{Q\in
\widetilde{\MM}^f_{\Omega,\phi_0}}E_{Q}[g]$. The equality with the supremum over $\MM_{\Omega,\phi_0}$ and $\widetilde{\MM}_{\Omega,\phi_0}$ follows from the same argument for the proof of Theorem 1.1, Step 2, in \cite{BFM16b}.

\end{prova}

\begin{prova}[Proof of Theorem \ref{thm: AB theorem 2}]

(3) $\Rightarrow$ (2) and (2) $\Rightarrow$ (1) are obvious.

\textbf{Step 1. }  To show that (1) implies (3). Suppose there is
no Uniformly Strong Arbitrage on $\Omega$ in
$\mathcal{A}_{\Phi}(\widetilde{\mathbb{F}})$. We first know from
Proposition \ref{prop: AB theorem 1} that $\MM_{\Omega,
\phi_0}\neq \emptyset$.  We can use a variational argument to
deduce the following equalities:  fix an arbitrary $K>0$ and let
$\widetilde{m}:\R^{d\times (T+1)}_+ \to \R$ defined as in
\eqref{eq: wt m}
\begin{align}
\pi_{\Omega^*, \Phi}(g(S)) =&\; \inf_{X\in \Lin(\Phi/\{\phi_0\})}\pi_{\Omega^*, \phi_0}(g(S) - X) \nonumber \\
=&\; \inf_{X\in \Lin(\Phi/\{\phi_0\})}\sup_{Q\in \MM_{\Omega, \phi_0}} E_{Q}[g(S) - X]  \label{eq: variational 1}\\
=&\; \inf_{X\in \Lin(\Phi/\{\phi_0\})}\sup_{Q\in \MM_{\Omega,
\phi_0}} E_{Q}[g(S) - X-K\widetilde{m}(S)]  \label{eq: variational 2}
\end{align}
where the second equality follows from Proposition \ref{prop: AB
theorem 1}.
Denote by $\mathcal{Q}$ (respectively $\widetilde{\mathcal{Q}}$)
the set of law of $S$ under the measures $ Q\in \MM_{\Omega,
\phi_0}$ (respectively $Q\in \widetilde{\MM}_{\Omega, \phi_0}$)
and write $\Lin(\{g_i\}_{i \in I/\{0\}})$ for the set of finite
linear combinations of elements in $\{g_i\}_{i \in I/\{0\}}$.
Observe that from Step 1 of the proof of Theorem 1.3 in
\citet{AB13} $\widetilde{\mathcal{Q}}$ is weakly compact and hence
the same is true for $\overline{\mathcal{Q}}$ (the weak closure of
$\mathcal{Q}$).

By a change of variable we have
$$\inf_{X\in \Lin(\Phi/\{\phi_0\})}\sup_{Q\in \MM_{\Omega, \phi_0}} E_{Q}[g(S) - X-K\widetilde{m}(S)]
= \inf_{G\in \Lin(\{g_i\}_{i \in I/\{0\}})}\sup_{\Q\in
\mathcal{Q}}\mathbb{E}_{\Q}[\widetilde{g}(\mathbb{S})],$$ where
$\mathbb{S} := (\mathbb{S}_t)_{t=0}^T$ is the canonical process on
$\R_+^{d\times (T+1)}$ and $\widetilde{g} = g - G -
K\widetilde{m}$. We aim at applying min--max theorem (see
Corollary 2 in \citet{T72}) to the compact convex set
$\overline{\mathcal{Q}}$, the convex set $\Lin(\{g_i\}_{i \in
I/\{0\}})$, and the function
\begin{align*}
f(\Q,G)=\int_{\R_+^{d \times (T+1)}} \Big(g(s_0,\ldots,s_T) -
G(s_0,\ldots,s_T)-K\widetilde{m}(s_0,\ldots,s_T)\Big) d \Q(s_0,\ldots,s_n).
\end{align*}
Clearly $f$ is affine in each of the variables. Furthermore, we
show that $f(\cdot, G)$ is upper  semi-continuous on
$\overline{\mathcal{Q}}$. To see this, fix $G\in \Lin(\{g_i\}_{i
\in I/\{0\}})$.  By definition of $f$ we have that
\begin{equation}
f(\Q,G) =  \mathbb{E}_{\Q}[\widetilde{g}(\mathbb{S})].
\end{equation}
It follows from Assumption \ref{ass:options} and \eqref{eq:
payoff_dominatedA} that $\widetilde{g}$ is bounded from above.
Hence, for every sequence of $\{\Q_n\}_n \in
\overline{\mathcal{Q}}$ with $\Q_n \to \Q$ as $n\to \infty$ for
some $\Q$ weakly, we have
\begin{equation*}
\lim_{n\to
\infty}\mathbb{E}_{\Q_n}[\widetilde{g}^+(\mathbb{S})]\leq \mathbb{E}_{\Q}[\widetilde{g}^+(\mathbb{S})]
\end{equation*}
by Portmanteau theorem, and
\begin{equation*}
\liminf_{n\to
\infty}\mathbb{E}_{\Q_n}[\widetilde{g}^-(\mathbb{S})] \ge
\mathbb{E}_{\Q}[\widetilde{g}^-(\mathbb{S})]
\end{equation*}
by Fatou's lemma, where $\widetilde{g} := \widetilde{g}^+ -
\widetilde{g}^-$ with $\widetilde{g}^+ := \max\{\widetilde{g}, 0\}$, $\widetilde{g}^-:=(-\widetilde{g})^+$.
Then
\begin{equation*}
\limsup_{n\to \infty}f(\Q_n, G) = \limsup_{n\to
\infty}\mathbb{E}_{\Q_n}[\widetilde{g}(\mathbb{S})]\le
\mathbb{E}_{\Q}[\widetilde{g}(\mathbb{S})] = f(\Q, G).
\end{equation*}

Therefore, the assumptions of Corollary 2 in \citet{T72} are
satisfied and we have, by recalling $\mathcal{Q}\subseteq
\overline{\mathcal{Q}}$ and equation \eqref{eq: variational 2},
\begin{align}
\pi_{\Omega^*, \Phi}(g(S)) =&\inf_{X\in \Lin(\Phi/\{\phi_0\})}\sup_{Q\in \MM_{\Omega,
\phi_0}}E_{Q}[g(S) - X -K\widetilde{m}(S)] \nonumber \\ =
&\, \inf_{G\in \Lin(\{g_i\}_{i \in I/\{0\}})}\sup_{\Q\in \mathcal{Q}}\mathbb{E}_{\Q}[g(\mathbb{S}) - G(\mathbb{S}) -K\widetilde{m}(\mathbb{S})] \nonumber\\
\leq &\,\inf_{G\in \Lin(\{g_i\}_{i \in I/\{0\}})}\sup_{\Q\in
\overline{\mathcal{Q}}}\mathbb{E}_{\Q}[g(\mathbb{S}) -
G(\mathbb{S}) -K\widetilde{m}(\mathbb{S})] \nonumber \\
= &\,\sup_{\Q\in \overline{\mathcal{Q}}}\inf_{G\in \Lin(\{g_i\}_{i
\in I/\{0\}})}\mathbb{E}_{\Q}[g(\mathbb{S}) -
G(\mathbb{S}) -K\widetilde{m}(\mathbb{S})] \nonumber\\
\leq &\,\sup_{\Q\in \widetilde{\mathcal{Q}}}\inf_{G\in \Lin(\{g_i\}_{i \in I/\{0\}})}
\mathbb{E}_{\Q}[g(\mathbb{S}) - G(\mathbb{S}) -K\widetilde{m}(\mathbb{S})] \nonumber \\
=&\, \sup_{Q\in \widetilde{\MM}_{\Omega, \phi_0}} \inf_{X\in
\Lin(\Phi/\{\phi_0\})} E_{Q}[g(S)- X -K\widetilde{m}(S)].\label{eq: minmax}
\end{align}

Take $g = 0$. If $\widetilde{\MM}_{\Omega, \Phi}= \emptyset$, then
$$ \sup_{Q\in \widetilde{\MM}_{\Omega, \phi_0}}\inf_{X\in \Lin(\Phi/\{\phi_0\})} E_{Q}[- X -K\widetilde{m}(S)] = -\infty, $$
and hence $\pi_{\Omega^*, \Phi}(0) = -\infty$, which contradicts
the no arbitrage assumption. Therefore we have (1) implies (3).

\textbf{Step 2. } To show the pricing--hedging duality, suppose
now $\widetilde{\MM}_{\Omega, \Phi}\neq \emptyset$. If $Q\not\in \widetilde{\MM}_{\Omega, \Phi}$, then
$$ \inf_{X\in \Lin(\Phi/\{\phi_0\})} E_{Q}[g(S)- X -K\widetilde{m}(S)] = - \infty. $$
Therefore, in \eqref{eq: minmax},
 it suffices to look at measures in
$\widetilde{\MM}_{\Omega, \Phi}\neq \emptyset$ only, and hence we
obtain \begin{eqnarray*}\pi_{\Omega^*, \Phi}(g(S)) & \leq &
\sup_{Q\in \widetilde{\MM}_{\Omega, \Phi}} E_{Q}[g(S)
-K\widetilde{m}(S)] \\& \leq & \sup_{Q\in \widetilde{\MM}_{\Omega,
\Phi}} E_{Q}[g(S)]+K \sup_{Q\in \widetilde{\MM}_{\Omega,
\Phi}}E_{Q}[-\widetilde{m}(S)]
\\ & = & \sup_{Q\in \widetilde{\MM}_{\Omega,
\Phi}} E_{Q}[g(S)]+K(T+1) \sup_{Q\in \widetilde{\MM}_{\Omega,
\Phi}}E_{Q}[-g_0(S)]
\end{eqnarray*}
Since $-g_0$ is bounded from above, the quantity $\sup_{Q\in
\widetilde{\MM}_{\Omega, \Phi}}E_{Q}[-g_0(S)]$  is finite and, by recalling that $K>0$ is arbitrary, we
get the thesis for $K\downarrow 0$.
\end{prova}

\section{Appendix}
    Let $X$ be a Polish space. The so-called \emph{projective hierarchy} (see \citet{Krechis} Chapter V) is constructed as follows. The first level is composed of the analytic sets $\Sigma^1_1$ (projections of closed subsets of $X\times{\N^\N}$), the co-analytic sets $\Pi^1_1$ (complementary of analytic sets), and the Borel sets $\Delta^1_1=\Sigma^1_1\cap \Pi^1_1$. The subsequent level are defined iteratively through the operations of projection and complementation. Namely,
\begin{eqnarray*}
    \Sigma^1_{n+1}&=& \text{projections of } \Pi^1_n \text{ subsets of } X\times\N^{\N}, \\
    \Pi^1_{n+1}&=& \text{complementary of sets in }\Sigma^1_{n+1},\\
    \Delta^1_{n+1}&=&\Sigma^1_{n+1}\cap \Pi^1_{n+1}.
\end{eqnarray*}
From the definition it is clear that $\Sigma^1_n\subseteq \Sigma^1_{n+1}$,
for any $n\in\N$, and analogous inclusions hold for $\Pi^1_n$ and $\Delta^1_n$.
Sets in the union of the projective classes (also called Lusin classes) are called
\emph{projective sets}, which we denoted by $\pr:=\bigcup_{n=1}^\infty\Delta^1_n=\bigcup_{n=1}^\infty\Sigma^1_n=\bigcup_{n=1}^\infty\Pi^1_n$.

\begin{remark} We observe that $\Sigma_1^1\cup \Pi^1_1$ is a sigma
algebra which actually coincides with $\mathcal{F}^{\mathcal{A}}$.
Moreover $\Sigma_1^1\cup
\Pi^1_1=\mathcal{F}^{\mathcal{A}}\subseteq \Delta_2^1$.
\end{remark}

\medskip

We first recall the following result from \citet{Krechis} (see Exercise 37.3).

\begin{lemma}\label{lem:img} Let $f:X\mapsto\mathbb{R}^k$ be Borel measurable. For any $n\in\N$,
\begin{enumerate}
 \item $f^{-1}(\Sigma^1_n)\subseteq\Sigma^1_n$;
 \item $f(\Sigma^1_{n})\subseteq\Sigma^1_{n}$.
\end{enumerate}
\end{lemma}

The following is a consequence of the previous Lemma.

\begin{lemma}
\label{Fsigma}
%
%
Let $f:X\mapsto\mathbb{R}^k$ be Borel measurable. For any $n\in\N$,
\begin{enumerate}
    \item $f^{-1}(\sigma(\Sigma^1_n))\subseteq\sigma(\Sigma^1_n)$;
    \item $f(\sigma(\Sigma^1_n))\subseteq\Sigma^1_{n+1}$.
\end{enumerate}
\end{lemma}

\begin{prova}
From Lemma \ref{lem:img} the first claim hold for
$\Sigma^1_n$ which generates the sigma-algebra. In particular, $\Sigma^1_n$ is contained in
\begin{eqnarray*}
    \left\{A\in\sigma(\Sigma^1_n)\mid f^{-1}(A)\in \sigma(\Sigma^1_n) \right\} &\subseteq & \sigma(\Sigma^1_n).
\end{eqnarray*}
Since the above set is a sigma-algebra, it also contains $\sigma(\Sigma^1_n)$, from which the claim follows. For the second assertion, we recall that $\Delta^1_n$ is a sigma-algebra for any $n\in\N$ (see Proposition 37.1 in \citet{Krechis}). In particular,
$$\sigma(\Sigma^1_n)\subseteq \sigma(\Sigma^1_n\cup\Pi^1_n)\subseteq \Delta^1_{n+1}\subseteq\Sigma^1_{n+1}.$$
Since, from Lemma \ref{lem:img}, $f(\Sigma^1_{n+1})\subseteq\Sigma^1_{n+1}$, the thesis follows.
\end{prova}

\begin{remark}\label{rk:projective_is_universal}
We recall that under the axiom of Projective Determinacy the class $\pr$, and hence also $\prF$, is included in the universal completion of $\mathcal{B}_X$ (see Theorem 38.17 in \cite{Krechis}). This axiom has been thoroughly studied in set theory and it is implied, for example, by the existence of infinitely many Woodin cardinals (see e.g. \cite{projective}).
\end{remark}

\subsection{Remark on conditional supports}\label{conditional:support}
Let $\mathcal{G}\subseteq \mathcal{B}_X$ be a countably generated
sub $\sigma$-algebra of $\mathcal{B}_X$. Then there exists a
proper regular conditional probability, i.e. a function
$\Prob_{\mathcal{G}}(\cdot ,\cdot ):(X ,\mathcal{B}_X)\mapsto
\lbrack 0,1]$ such that:

\begin{itemize}
\item[a)] for all $\omega \in \Omega $,
$\Prob_{\mathcal{G}}(\omega ,\cdot )$ is a probability measure on
$\mathcal{B}_X$;

\item[b)] for each (fixed) $B\in \mathcal{B}_X$, the function
$\Prob_{\mathcal{G}}(\cdot ,B)$ is $\mathcal{G}$-measurable and a
version of $E_{\Prob}[\mathbf{1}_B\mid \mathcal{G}](\cdot)$ (here
the null set where they differ depends on $B$);

\item[c)] $\exists N\in \mathcal{G}$ with $\Prob(N)=0$ such that $\Prob_{\mathcal{G}%
}(\omega ,B)=\mathbf{1 }_{B}(\omega )$ for $\omega \in X \setminus
N$ and $B\in \mathcal{G}$ (here the null set where they differ
does not depend on $B$); moreover for all $\omega \in X \setminus
N$, we have $\Prob_{\mathcal{G}}(\omega,A_{\omega})=1$ where
$A_{\omega}=\bigcap \{A:\omega\in A,\,A\in
\mathcal{G}\}\in\mathcal{G}.$

\item[d)] For every $A\in\mathcal{G}$ and $B\in\mathcal{B}_X$ we
have $\Prob(A\cap B)=\int_A
\Prob_{\mathcal{G}}(\omega,B)\Prob(d\omega) $
\end{itemize}

We now consider a measurable $\xi:X\to \R^d$ and $P_{\xi}:X\times
\mathcal{B}_{\R^d}\rightarrow [0,1]$ defined by
$$P_{\xi}(\omega,B):=\Prob_{\mathcal{G}}(\omega,\{\tilde\omega\in A_{\omega}\mid \xi(\tilde{\omega})\in B\}),$$
and observe that from a) and with $N$ as in c), for any $\omega\in X\setminus N$,
$P_{\xi}(\omega,\cdot)$ is a probability measure on
$(\R^d,\mathcal{B}_{\R^d})$. Finally, we let $B_{\varepsilon}(x)$ denote the ball of radius
$\varepsilon$ with center in $x$, and we introduce the closed
valued random set
\begin{equation}\label{condSupp}
\omega\to \chi_{\mathcal{G}}(\omega):=\{x\in \R^d \mid
P_{\xi}(\omega,B_{\varepsilon}(x))>0\;\forall\, \varepsilon>0 \},
\end{equation}
for $\omega\in X\setminus N$ and $\R^d$ otherwise.  $\chi_{\mathcal{G}}$ is
$\mathcal{G}$-measurable since, for any open set $O\subseteq\R^d$,
we have $$\{\omega\in X\mid \chi_{\mathcal{G}}(\omega)\cap
O\neq\emptyset\}=N\cup \{\omega\in X\setminus N\mid
P_{\xi}(\omega,O)>0\}=N\cup\{\omega\in X\setminus N\mid
\Prob_{\mathcal{G}}(\omega,\xi^{-1}(O)\cap A_\omega)>0\},$$ with
the latter belonging to $\mathcal{G}$ from b) and c) above. By
definition $\chi_{\mathcal{G}}(\omega)$ is the support of
$P_{\xi}(\omega,\cdot)$ and therefore, for every $\omega\in X$,
$P_{\xi}(\omega,\chi_{\mathcal{G}}(\omega))=1$. Notice that since
the map $\chi_{\mathcal{G}}$ is $\mathcal{G}$-measurable then for
 $\omega\in X$ we have
$\chi_{\mathcal{G}}(\omega)=\chi_{\mathcal{G}}(\tilde\omega)$ for
all $\tilde\omega\in A_{\omega}$.

\begin{lemma}\label{probSupp} Under the previous assumption we have $\{\omega\in X\mid \xi(\omega) \in \chi_{\mathcal{G}}(\omega)\}\in\mathcal{B}_X$ and $\Prob(\{\omega\in X\mid \xi(\omega) \in
\chi_{\mathcal{G}}(\omega)\})=1$.
\end{lemma}

\begin{prova} Set $B:=\{\omega\in X\mid \xi(\omega) \in \chi_{\mathcal{G}}(\omega)\}$. $B\in\mathcal{B}_X$ follows from the measurability of $\xi$ and $\chi_{\mathcal{G}}$. From the properties of regular conditional
probability we have
$$\Prob(B)=\Prob(B\cap X)= \int_{X}
\Prob_{\mathcal{G}}(\omega,B)\Prob(d\omega).$$ Consider the atom
$A_{\omega}=\cap \{A:\omega\in A,\,A\in \mathcal{G}\}.$ From
property c) we have $\Prob_{\mathcal{G}}(\omega,A_{\omega})=1$ for
any $\omega\in X\setminus N$. Therefore for every $\omega\in
X\setminus N$ we deduce
\begin{eqnarray*}\Prob_{\mathcal{G}}(\omega,B)& = &\Prob_{\mathcal{G}}(\omega,B\cap
A_{\omega})=\Prob_{\mathcal{G}}(\omega,\{\tilde\omega\in
A_{\omega}\mid \xi(\tilde{\omega})\in
\chi_{\mathcal{G}}(\tilde\omega)\})
\\ & = & \Prob_{\mathcal{G}}(\omega,\{\tilde\omega\in
A_{\omega}\mid \xi(\tilde{\omega})\in
\chi_{\mathcal{G}}(\omega)\})=P_{\xi}(\omega,\chi_{\mathcal{G}}(\omega))=1.
\end{eqnarray*}
Therefore
$$\Prob(B)=\int_{X}
\Prob_{\mathcal{G}}(\omega,B)\Prob(d\omega)=\int_{X}
\mathbf{1}_{X\setminus N}(\omega)\Prob(d\omega)=1.$$
\end{prova}

\section*{Acknowledgments.}
Zhaoxu Hou gratefully acknowledges the support of the Oxford-Man Institute of Quantitative Finance. Jan Ob\l\'oj gratefully acknowledges funding received from the European Research Council under the European Union's Seventh Framework Programme (FP7/2007-2013) / ERC grant agreement no. 335421. Jan Ob\l\'oj is also thankful to the Oxford-Man Institute of Quantitative Finance and St John's College in Oxford for their financial support.

\bibliographystyle{apalike}

\bibliography{RTAP}

\end{document}